\documentclass[nofootinbib,aps,11pt,preprintnumbers,unsortedaddress,prd]{revtex4-1}
\usepackage{graphicx}
\usepackage{cancel}
\usepackage{amssymb}
\usepackage{textcomp}
\usepackage{amsmath}
\usepackage{bm}
\usepackage{times}
\usepackage{epsfig}
\usepackage{color}
\usepackage{graphics}
\usepackage{hyperref}
\usepackage{setspace}
\usepackage{epsfig}
\usepackage{amsmath}
\usepackage{bm}
\usepackage{times}
\usepackage{graphicx}
\usepackage{color}
\usepackage{slashed}
\usepackage{graphicx}
\usepackage{amsmath, amssymb}

\hypersetup{
    pdfnewwindow=true,      
    colorlinks=true,       
    linkcolor=black,          
    citecolor=blue,        
    filecolor=blue,      
    urlcolor=blue           
}
\def\nn{\nonumber}
\def\bea{\begin{eqnarray}}
\def\eea{\end{eqnarray}}
\begin{document}
\title{\Large {\it{\bf{On Higgs Decays, Baryon Number Violation, and SUSY at the LHC}}}}
\author{Jonathan M. Arnold}
\affiliation{California Institute of Technology, Pasadena, CA 91125, USA}
\author{Pavel Fileviez P\'erez}
\affiliation{Center for Cosmology and Particle Physics (CCPP) \\
New York University, 4 Washington Place, NY 10003, USA }
\author{Bartosz Fornal}
\affiliation{California Institute of Technology, Pasadena, CA 91125, USA}
\author{Sogee Spinner}
\affiliation{International School for Advanced Studies (SISSA), \\
Via Bonomea 265, 34136 Trieste, Italy \\}
%
\date{\today}
\vspace{1.0cm}
\begin{abstract}
Baryon number violating interactions could modify the signatures of supersymmetric models at the Large Hadron Collider.
In this article we investigate the predictions for the Higgs mass and the Higgs decays in a simple extension 
of the minimal supersymmetric standard model where the local baryon and lepton numbers are spontaneously 
broken at the TeV scale. This theory predicts baryon number violation at the low scale which can change 
the current LHC bounds on the supersymmetric spectrum. Using the ATLAS and CMS bounds on the 
Higgs mass we show the constraints on the sfermion masses, and show the subsequent predictions for the radiative Higgs decays.
We found that the Higgs decay into two photons is suppressed due to the existence of new light leptons.
In this theory the stops can be very light in agreement with all experimental bounds and we 
make a brief discussion of the possible signals at the LHC.
\end{abstract}
\maketitle
%
\newpage
\begin{spacing}{2.0}
\tableofcontents
\end{spacing}
\newpage
\section{Introduction}
The minimal supersymmetric extension of the standard model (MSSM) is considered as one of the most appealing theories 
to describe physics at the TeV scale. While this theory makes many interesting predictions the signals at the Large Hadron 
Collider (LHC) depend on unknowns such as the supersymmetric spectrum and the presence or absence of baryon (B) and lepton (L) number violating interactions (collectively known as R-parity violating interactions). The CMS~\cite{SUSY-CMS} and ATLAS~\cite{SUSY-ATLAS} experiments have studied 
many possible signals of the MSSM at the LHC setting very strong bounds on the gluino and squarks masses in 
some specific scenarios with missing energy. In the majority of the experimental studies it is assumed the absence of the baryon (B) and 
lepton (L) number violating interactions (R-parity conservation). However, it is well-known that in general the B and L symmetries can be 
broken changing many of the predictions for the collider experiments. For example, one could modify all collider bounds 
based on the searches for missing energy if the baryon number is broken. 

In this article we discuss the possible impact of the baryon number violating interactions on the supersymmetric 
signals at the LHC. We focus our study in the context of a simple extension of the MSSM where the baryon and 
lepton numbers are local gauge symmetries spontaneously broken at the TeV scale. We refer to this theory as the ``BLMSSM"~\cite{BLMSSM}. 
The main motivation for this theory is that a large desert between 
the electroweak scale and grand unified scale is no longer necessary since, while B and L are broken at the low scale the  proton remains stable.
In the BLMSSM the lepton number is broken in an even number while the baryon number violating operators 
can change B by one unit. Even though new generations of fermions are required, they do not mix with the SM fermions and therefore do not lead to flavour violation at tree level. Furthermore, they are not associated with Landau poles at the low scale. The light Higgs boson mass 
can be large without assuming a large stop mass and left-right mixing and one could modify the current 
LHC bounds on the supersymmetric spectrum due to the presence of the baryon number violating interactions.

We study in great detail the correlation between the Higgs mass and the decay of the Higgs boson into 
two photons following the new results presented by the ATLAS and CMS collaborations. In this theory the new light leptons 
appreciably decrease the predictions for the Higgs decay into two gammas. Therefore, confirmation of the two photon signal at the LHC as the Higgs decay rules out this model. In this theory the stops can be very light in agreement with the Higgs mass and colliders bounds.

This article is organized as follows: In section II we discuss the main features of the BLMSSM. 
The possible impact of the baryon number violating interactions on the LHC searches are 
discussed in section III. The predictions for the light CP-even Higgs mass and the constraints 
on the supersymmetric spectrum are investigated in section IV. In section V the radiative Higgs 
decays are studied, while the evolution of the gauge and Yukawa couplings are investigated 
in section VI. In the appendices we include all details needed for the numerical calculations. 
\section{The BLMSSM}
In this article we study a simple supersymmetric model where the baryon number
(B) and lepton number (L) are local gauge symmetries~\cite{BLMSSM}.
This model is based on the gauge symmetry
$$G_{BL}=SU(3)_C \bigotimes SU(2)_L \bigotimes U(1)_Y \bigotimes U(1)_B \bigotimes U(1)_L$$
We refer to this model as the ``BLMSSM". In this context we have found that~\cite{BLMSSM}

\begin{itemize}

\item  The local B and L are spontaneously broken at the TeV scale.

\item There are no dangerous operators mediating proton decay.

\item The lepton number is broken in an even number while the baryon number violating operators can change B by one unit.

\item Anomaly cancellation requires the presence of new families, however there is no flavour violation at tree level since they do not mix with the SM fermions.

\item There are no Landau poles at the low scale due to the new families.

\item The light Higgs boson mass can be large without assuming a large stop mass and left-right mixing.

\item One could modify the current LHC bounds on the supersymmetric spectrum due to the presence of the baryon number violating interactions.

\end{itemize}
In this model we have the chiral superfields of the MSSM, and in order to cancel the B and L anomalies we need a vector-like family:
$\hat{Q}_4$, $\hat{u}_4^c$ ,  $\hat{d}_4^c$, $\hat{L}_4$,  $\hat{e}_4^c$,  $\hat{\nu}^c_4$ and $\hat{Q}_5^c$, $\hat{u}_5$, $\hat{d}_5$,
$\hat{L}_5^c$, $\hat{e}_5$, $\hat{\nu}_5$. See Table I for the superfields present in the BLMSSM.

\begin{flushleft}
\underline{Interactions:} The full superpotential of the model is given by
\end{flushleft}
\begin{equation}
{\cal W}_{\rm{BL}}={\cal W}_{\rm{MSSM}} \ + \  {\cal W}_{\rm{B}} \ + \  {\cal W}_{\rm{L}} \ + \ {\cal W}_{X} \ + \ {\cal W}_{5},
\end{equation}
where
\begin{equation}
{\cal W}_{\rm{MSSM}}=Y_u \hat{Q} \hat{H}_u \hat{u}^c \ + \ Y_d \hat{Q} \hat{H}_d \hat{d}^c \ + \ Y_e \hat{L} \hat{H}_d \hat{e}^c \ + \ \mu \hat{H}_u \hat{H}_d,
\end{equation}
is the MSSM superpotential and
\begin{eqnarray}
{\cal W}_{\rm{B}}&=&\lambda_{Q}  \hat{Q}_4  \hat{Q}_5^c \hat{S}_B \ + \  \lambda_{u}  \hat{u}_4^c  \hat{u}_5 \hat{\bar{S}}_B \ + \  \lambda_{d}  \hat{d}_4^c  \hat{d}_5 \hat{\bar{S}}_B \ + \   \mu_{B} \hat{\overline{S}}_B \hat{S}_B \nonumber \\
&+& Y_{u_4}  \hat{Q}_4 \hat{H}_u \hat{u}^c_4 \ + \ Y_{d_4} \hat{Q}_4 \hat{H}_d \hat{d}_4^c \ + \ Y_{u_5}  \hat{Q}^c_5 \hat{H}_d \hat{u}_5 \ + \ Y_{d_5} \hat{Q}_5^c \hat{H}_u \hat{d}_5.
\end{eqnarray}
The new quark superfields acquire TeV scale masses once the $S_B$ and $\bar S_B$ Higgs fields acquire a VEV. Consequently, the Yukawa couplings of the new quarks to the Higgs fields do not contribute greatly to the new quark masses and can be neglected. Furthermore the Yukawa couplings between the new quarks and the MSSM Higgses can be large and modify the Higgs mass at one-loop level. Notice that these couplings can have a large impact on the production cross section, $gg \to h$, making it difficult to satisfy the experimental bounds on Higgs production. In this work, we will take the conservative approach and assume these new quark Yukawa couplings are small. The Higgs mass is therefore only substantially modified by the Yukawa couplings of the new leptons which must be large to insure the new leptons masses are large enough to satisfy collider bounds.

In the leptonic sector one has the following interactions
\begin{eqnarray}
{\cal W}_{\rm{{L}} }&=& Y_{e_{4}}  \  \hat{L}_4 \hat{H}_d \hat{e}^c_4 \ + \  Y_{e_{5}}  \  \hat{L}_5^c \hat{H}_u \hat{e}_5 \ + \ Y_{\nu_{4}} \  \hat{L}_4 \hat{H}_u \hat{\nu}^c_4
                         +  Y_{\nu_{5}}  \  \hat{L}_5^c \hat{H}_d \hat{\nu}_5 \nonumber \\
                          &+&   Y_{\nu}  \  \hat{L}  \hat{H}_u \hat{\nu}^c \ + \  \lambda_{\nu^c}  \  \hat{\nu}^c  \hat{\nu}^c  \hat{\overline{S}}_L \ + \  \mu_{L} \hat{\overline{S}}_L \hat{S}_L.
                          \label{Wlept}
\end{eqnarray}
Notice that we have an implementation of the seesaw mechanism for the light neutrino masses once the $\bar S_L$ field acquires a VEV, while the new neutrinos have Dirac mass terms.
In order to avoid stability for the new quarks we add the fields, $\hat{X}$ and $\hat{\bar{X}}$, which have the following interactions
\begin{eqnarray}
{\cal W}_{\rm{X}}&=&\lambda_{1}  \hat{Q}  \hat{Q}_5^c \hat{X} \ + \  \lambda_{2}  \hat{u}^c  \hat{u}_5 \hat{\bar{X}} \ + \  \lambda_{3}  \hat{d}^c  \hat{d}_5 \hat{\bar{X}} \ + \  \mu_{X} \hat{\overline{X}} \hat{X},
\end{eqnarray}
where the baryon number for the new fields are: $B_X=2/3 + B_4=-B_{\bar{X}}$, and if we assume that they do not get a vev the lightest one can be a dark matter candidate even if R-parity is violated. See Refs.~\cite{FileviezPerez:2010gw} and~\cite{Dulaney:2010dj} for the use of this idea in a previous version of the model.

For any value of the baryonic charges of the new fermions, which satisfy the anomaly conditions, the Higgses $\hat{S}_B$ and
$\hat{\overline{S}}_B$ have charges $1$ and $-1$, respectively. Then, one can write the following dimension five operator which gives rise to baryon number violation once baryon number is broken through the VEV of $S_B$:
\begin{eqnarray}
\label{bnv}
{\cal W}_5 &=& \frac{a_1}{\Lambda} \hat{u}^c \hat{d}^c \hat{d}^c \hat{S}_B.
\end{eqnarray}
Therefore, after breaking $U(1)_B$ we find the so-called $\lambda^{''}$ MSSM interactions which can modify the current LHC bounds on the supersymmetric mass spectrum. Regardless of this R-parity breaking term, the field $X(\bar{X})$ or their superpartners can be a dark matter candidate.
The study of the properties of the dark matter candidates is beyond the scope of this article. 
\vspace{-1.1cm}
\subsection{B and L Symmetry Breaking}
In this section we discuss how the local B and L symmetries are broken. In this context the local gauge group, $G_{BL}$, is broken to $G_{SM} \bigotimes M_L$. 
Here $G_{SM}$ is the SM gauge group and $M_L=(-1)^L$ is the lepton parity:
\begin{displaymath}
G_{BL} \  \Longrightarrow \ G_{SM} \bigotimes M_L.
\end{displaymath}
In this model the local baryonic symmetry $U(1)_B$ is broken by the vev of the scalar fields $S_B \sim (1,1,0,1,0)$ and $\bar{S}_B \sim (1,1,0,-1,0)$.
The relevant scalar potential is given by
\begin{eqnarray}
V_B&=&\frac{g_B^2}{8} \left(  |S_B|^2 - |\bar{S}_B|^2 \right)^2 \ + \ \left(  |\mu_B|^2 + m_{S_B}^2 \right) |S_B|^2 \ + \  \left(  |\mu_B|^2 + m_{\bar{S}_B}^2 \right) |\bar{S}_B|^2 \nonumber \\
&-& \left( b_B  S_B \bar{S}_B \ + \  \rm{h.c.}\right).
\end{eqnarray}
Defining the vevs,  $\left< S_B\right>={v_B}/{\sqrt{2}}$ and $\left< \bar{S}_B\right>= {\bar{v}_B}/{\sqrt{2}}$, the minimization conditions read as
\begin{eqnarray}
|\mu_B|^2 + m_{S_B}^2 + \frac{1}{2} M_{Z_B}^2 \cos 2 \beta_B - b_B \tan \beta_B&=&0, \\
|\mu_B|^2 + m_{\bar{S}_B}^2 - \frac{1}{2} M_{Z_B}^2 \cos 2 \beta_B - b_B \cot \beta_B&=&0,
\end{eqnarray}
where $\tan \beta_B = \bar v_B/v_B$ and $M_{Z_B}^2= g_B^2 (v_B^2 + \bar{v}_B^2)/4$. Assuming that the potential is bounded from below along the D-flat direction
one finds the condition:
\begin{equation}
2 b_B < 2 |\mu_B|^2 + m_{S_B}^2 + m_{\bar{S}_B}^2,
\end{equation}
while
\begin{equation}
b_B^2 > \left(  |\mu_B|^2 +  m_{S_B}^2 \right) \left(  |\mu_B|^2 +  m_{\bar{S}_B}^2  \right),
\end{equation}
in order to have a non-trivial minimum. It is important to emphasize that the symmetry $U(1)_B$ is broken at the TeV scale and therefore the mass of the new neutral gauge boson is related
to the SUSY breaking mass scale. In order to show this we give the dependence of the new gauge boson masses on the  parameters in the model:
\begin{equation}
	\frac{1}{2} M_{Z_B}^2 = - |\mu_B|^2 \ + \ \left(  \frac{m_{S_B}^2  - m_{\bar{S}_B}^2 \tan^2 \beta_B}{\tan^2 \beta_B -1} \right).
\end{equation}
Once B is broken we can find new interactions which violate baryon number. Using Eq.(6) one finds
\begin{equation}
2  \lambda_{ijk}^{''}  \ u^c_i \ d_j^c \ \tilde{d}^c_k  \  \  {\rm{and}}  \   \  \lambda_{ijk}^{''}  \ \tilde{u}^c_i \ d_j^c \  d^c_k,
\end{equation}
with
\begin{equation}
\lambda_{ijk}^{''} = a_1^{ijk} \frac{v_B}{\Lambda \sqrt{2}},
\end{equation}
where $a_1^{ijk}=-a_1^{ikj}$. These interactions break baryon number in one unit and are the so-called $\lambda_{ijk}^{''}$ terms of the MSSM.

As in the case of the baryon number the local $U(1)_L$ is broken at the TeV scale by the vev of the scalar fields $S_L$ and $\bar{S}_L$.
Following the discussion above one can find a similar relation between the quark-phobic gauge boson mass and the soft terms of the scalar fields:
\begin{equation}
	\frac{1}{2} M_{Z_L}^2 = - |\mu_L|^2 \ + \ \left(  \frac{m_{S_L}^2 - m_{\bar{S}_L}^2 \tan^2 \beta_L}{ \tan^2 \beta_L - 1 } \right).
\end{equation}
In the above equations $m_{S_L} (m_{S_B})$ and $m_{\bar{S}_L} (m_{\bar{S}_B})$ are soft masses for the Higgses $S_L (S_B)$ and $\bar{S}_L (\bar{S}_B)$,
while $\tan \beta_L = \left< \bar S_L\right> / \left< S_L\right>=\bar v_L/v_L$ and $M_{Z_L}^2= g_L^2 (v_L^2 + \bar{v}_L^2)$.

Now, after symmetry breaking one can see that the lepton number is only broken in two units. Using the term, $\hat{\nu}^c \hat{\nu}^c \hat{\bar{S}}_L$, 
in Eq. (4) one finds the Majorana mass term for the right-handed neutrinos
\begin{equation}
	\lambda_{\nu^c}^{ij}  \ \nu^c_i \  \nu^c_j \  \frac{\bar{v}_L}{\sqrt{2}},
\end{equation}
which is needed to generate neutrino masses though the seesaw mechanism.

Summarizing, in this model the local B and L symmetries can be broken at the TeV scale while contributions
to proton decay are not induced because the baryon number is broken by one unit and the lepton number is broken by two units
as required by the seesaw mechanism. There are relevant constraints coming from the $\Delta B=2$ processes, 
such as $n-\bar{n}$ oscillations and di-nucleon decays, and from cosmology which we will discuss in the next sections.
\subsection{Mass Spectrum}
\begin{flushleft}
\underline{New Leptons:} The mass for the new leptons are given by
\end{flushleft}
\begin{eqnarray}
M_{e_4}&=&Y_{e_4} \frac{v_d}{\sqrt{2}}, \  \ 
M_{e_5} =  Y_{e_5} \frac{v_u}{\sqrt{2}}, \ \ M_{\nu_4}=Y_{\nu_4} \frac{v_u}{\sqrt{2}},
\ 
{\rm{and}}
\
M_{\nu_5} =  Y_{\nu_5} \frac{v_d}{\sqrt{2}}.
\end{eqnarray}
Using the above equations and imposing the perturbative condition for the Yukawa coupling,
$Y_{e_4}^2/4 \pi \leq 1$, one finds an upper bound on $\tan \beta$: $$\tan \beta \leq 6.1 (4.3),$$ when $M_{e_4} \geq 100 (140)$ GeV.
\begin{flushleft}
%
\underline{Sfermion Masses}:
In this model the sfermion masses are modified due to existence of new D-terms. For example, the stop mass matrix reads as
\end{flushleft}
\begin{eqnarray}
&&
\left(
\begin{array}{cc}
      M_t^2 + M_{\tilde{Q}_3}^2 \ + \ \left( \frac{1}{2} - \frac{2}{3}  \sin^2 \theta_W \right) M_Z^2 \cos 2 \beta +  \frac{1}{3} D_B
	&
	M_t  \ X_t
	\\
	M_t \ X_t
	&
	M_t^2 \ + \ M_{\tilde{u}_3^c}^2 + \frac{2}{3} \sin^2 \theta_W M_Z^2 \cos 2 \beta  -  \frac{1}{3} D_B
\end{array}
\right), \nonumber \\
\end{eqnarray}
where $M_{\tilde{Q}_3}^2$ and $M_{\tilde{u}_3^c}^2$ are squark soft masses. $D_B=\frac{1}{2} M_{Z_B}^2 \cos 2 \beta_B$ define the new
contribution due to the presence of the $U(1)_B$ D-term, $M_t$ is the top mass and $X_t = A_t - \mu \cot \beta$ is the left-right mixing in the stop sector. 
In a similar way we can write the mass matrix for the sbottoms. See the Appendix for details.

The new slepton mass are modified by the $U(1)_L$ D-term. In the case of the mass matrix for the new sleptons of the 
fifth generation reads as
\begin{eqnarray}
&&
{\cal M}_{\tilde{\nu}_5}^2=
\left(
\begin{array}{cc}
      M_{\nu_5}^2 + M_{\tilde{L}_5^c}^2 \ - \  \frac{1}{2} M_Z^2 \cos 2 \beta -(3+L_4) D_L
	&
	M_{\nu_5}  \ X_{\nu_5}
	\\
	M_{\nu_5} \ X_{\nu_5}
	&
	M_{\nu_5}^2 \ + \ M_{\tilde{\nu}_5}^2 + (3+L_4) D_L
\end{array}
\right).
\end{eqnarray}
Here $D_L= -\frac{1}{4} M_{Z_L}^2 \cos 2 \beta_L$ is the new D-term contribution and $X_{\nu_5}=A_{\nu_5} - \mu \tan \beta$ the left-right mixing in this sector.
In order to simplify the discussion in the text we list the mass matrices for the other sfermions in the Appendix.
\section{Baryon Number Violation and the SUSY Spectrum}
When the baryon asymmetry is generated above the electroweak scale strong bounds on the $\lambda^{''}_{ijk} u^c_i d^c_j d^c_k$ couplings exist from the condition
that the $2 \to 2$ and $2 \to 1$ processes do not washout the baryon asymmetry generated. These constraints have been studied in great details in
Refs.~ \cite{Campbell:1990fa, Campbell:1991at, Fischler:1990gn, Dreiner:1992vm}. The bound on these couplings read as~\cite{Campbell:1991at,Dreiner:1992vm}
\begin{equation}
\lambda^{''}_{ijk} \lesssim 5 \times 10^{-7} \left(  \frac{M_{\tilde q}}{ 1 \rm{TeV} } \right)^{1/2},
\end{equation}
where $M_{\tilde q}$ is the squark mass. Now, in order to understand the impact of this bound on the SUSY signals we will consider different scenarios for the LSP:
\begin{itemize}
\item Neutralino as the LSP: 

In this case the neutralino will decay into three quarks and one can have the following signals
$$pp \ \to \ \tilde{t}^* \tilde{t} \  \to \bar{t} t  \ \tilde{\chi}_1^0  \tilde{\chi}_1^0 \  \to \   \bar{t} t  \ 6 j, \  \  \ pp \ \to \ \tilde{b}^* \tilde{b} \  \to \bar{b} b  \ \tilde{\chi}_1^0  \tilde{\chi}_1^0 \  \to \   \bar{b} b  \ 6 j .$$
Therefore, one could modify the bounds on the supersymmetric spectra since there is no missing energy in these channels. 
The neutralino decay length can naively be estimated as
\begin{equation}
	\text{L} (\tilde{\chi}^0 \to 3 q) \ > \  160 \ \text{m} \left(\frac{M_{\tilde q}}{500 \text{ GeV}}\right)^4 
	\left(\frac{100 \text{ GeV}}{M_{\tilde \chi^0}}\right)^5 \left(\frac{2.5 \times 10^{-7}}{\lambda''}\right)^2,
\end{equation}
assuming the cosmological bounds. Therefore, the lightest neutralino would decay outside the detector and one has the standard signals with missing energy at the LHC. However, if the baryogenesis 
mechanism is at the weak scale one can avoid the bounds from cosmology and the neutralino can decay inside the detector.

\item Gluino as the LSP: 

In this case the gluino pair production can lead to channels with same-sign tops and multijets
$$pp \ \to \ \tilde{g} \tilde{g} \  \to t t  \ 4 j , b b 4 j.$$
Now, assuming the constraint coming from cosmology one can estimate naively the decay length of the gluino as
\begin{equation}
	\text{L} (\tilde{g} \to 3 q) \ > \  10 \text{ m} \left(\frac{M_{\tilde q}}{10^3 \text{ GeV}}\right)^4 
	\left(\frac{400 \text{ GeV}}{M_{\tilde g}}\right)^5 \left(\frac{10^{-7}}{\lambda''}\right)^2.
\end{equation}
Therefore, the gluino is long-lived and form bounded states. The resulting states that consist of either of a gluino pair or triplets of quarks, or of a gluino bound to a gluon, are called R-hadrons~\cite{Farrar:1978xj}. If the
gluinos are produced near threshold, the formation of gluino-pair bound states
(gluinonium) is also possible and leads to characteristic signals~\cite{Kuhn:1983sc, Goldman:1984mj, Bigi:1991mi, Chikovani:1996bk, Cheung:2004ad} and place strong bounds on the gluino mass.

\item Slepton as the LSP: 

If the LSP is a charged selectron one has a long-lived charged track since the decay length is very large due to the bound coming from cosmology
and the four-body phase space suppression. This scenario is very similar to the long-lived stau scenario in gauge mediation~\cite{Drees:1990yw, Feng:1997zr} 
and one can have signals with two leptons, a same-sign top pair and four jets
$$pp \ \to \ \tilde{e}^*_i  \tilde{e}_i \  \to e^+_i e^-_i \ t t  \ 4 j, \ e^+_i e^-_i \ b b \ 4 j.$$
In the case when the sneutrino is the LSP one has missing energy and multijets 
$$pp \ \to \ \tilde{\nu}^*_i  \tilde{\nu}_i \  \to \bar{\nu} \nu \ t t  \ 4 j, \ \bar{\nu} \nu \ b b  \ 4 j.$$
This scenario is possible and will have constraints from the missing energy searches.

\item Chargino as the LSP: 

The case of a long-lived chargino is very similar to the previous scenario where one has a long-lived charged slepton. One has a charged
tracks due to existence of a long-lived charged particle and we can have the following signals
$$pp \ \to \ \tilde{\chi}^+_i  \tilde{\chi}^-_i \  \to W^+ W^-  \ t t  \ 4 j, \ W^+ W^-  \ b b  \ 4 j,$$
with same-sign top pair and multijets.

\item Squark as the LSP: 

Due to the cosmological constraint the squarks will be long-lived and form bounded states. If we compute the decay length of a squark one finds
\begin{equation}
\text{L} (\tilde{q}_i \to q_j q_k) \ > \  1 \ \text{mm} \left(\frac{10^2 \ \text{GeV}} {M_{\tilde q}} \right) \left(\frac{10^{-7}}{\lambda''}\right)^2.
\end{equation}
Therefore, the squark will form bounded states but it will decay inside the detector.
In this case we can have displaced vertices as well when the stop (sbottom) has mass around 100 GeV and we can have signals with four jets
$$pp \ \to \ \tilde{t}^*  \tilde{t} \  \to \  4 j, \  pp \ \to \ \tilde{b}^*  \tilde{b} \  \to \  4 j.$$
Therefore, one can avoid the LHC constraints coming from the searches for multijets and missing energy.
Notice that this scenario is quite relevant for us because we will study different cases where the stop is quite light.
\end{itemize}
It is important to mention that in the model discussed in this article the above constraints are relevant even if baryon number 
is broken at the TeV scale. In general one can have an asymmetry in the SM sector and in the dark matter sector, where 
we have the $X$ field. Now, we have discussed above that the $u^c d^c d^c$ interactions are generated at the TeV scale, 
and if they are in thermal equilibrium before the electroweak phase transition, around 100 GeV, one cannot preserve the 
baryon asymmetry in the visible sector. Therefore, in order to make sure that the asymmetry in the visible sector 
is not washed out we impose the above constraints. 

It is well-known, that if the baryon asymmetry is generated below the electroweak scale the bounds on $\lambda^{''}_{ijk}$ listed above are not present. 
However, there are other bounds on these couplings. The most important coming from dinucleon decay, $pp \to K^{+} K^{-}$~\cite{Litos}, and one gets
$\lambda^{''}_{uds} < 10^{-8}$~\cite{Goity:1994dq}. If we use this bound and the one above from cosmology we can imposse a lower bound on the cutoff of the theory.
Using $\lambda^{''} < 10^{-8}$, $\left< S_B \right> \sim 1$ TeV and $a_1 \sim 1$ one gets 
\begin{equation}
\frac{a_1}{\Lambda} \  \left< S_B \right> \ < \ 10^{-8} \  \Longrightarrow  \Lambda > 10^{11} \  {\rm{GeV}}.
\end{equation}
This is the naive lower bound on the cutoff of the theory. Of course, the coupling $a_1$ can be smaller and the cutoff of the theory can be much lower.
In the last section we will use the running of the Yukawa couplings to set the possible cutoff assuming perturbativity at the high scale.
\section{The Light CP-even Higgs Mass}
Recently, the ATLAS collaboration has published a new combined analysis~\cite{Collaboration:2012sk,Collaboration:2012si} which excludes 
a SM Higgs with mass in the ranges 112.9 GeV-115.5 GeV, 131 GeV-238 GeV, and 251 GeV-466 GeV.  While the 
CMS collaboration excludes a SM Higgs with mass in the range 127.5 GeV-600 GeV~\cite{Collaboration:2012tx,Collaboration:2012tw,CMSnew}. 
Also, it is well-known that both collaborations have observed an excess around 125 GeV. 

In this article we will consider a conservative scenario where the light CP-even Higgs is SM-like with mass 
in the range 115-128 GeV, and using this range we will show the possible constraints on the supersymmetric 
spectra in the MSSM and in the BLMSSM. In this way we can compare both models and predictions for the 
radiative Higgs decays showing the possibility of ruling out the BLMMSM if the excess around 
125 GeV is confirmed in the new analysis by the ATLAS and CMS collaborations.   

In order to set our notation we define the neutral Higgses as
\begin{equation}
H_u^0=\frac{1}{\sqrt{2}} \left( v_u \ + \ h_u \right) + \frac{i}{\sqrt{2}} A_u,
\end{equation}
and
\begin{equation}
H_d^0=\frac{1}{\sqrt{2}} \left( v_d \ + \ h_d \right) + \frac{i}{\sqrt{2} }A_d.
\end{equation}
Using this notation and working in the basis $( h_d, h_u)$ the mass matrix for the MSSM neutral CP-even Higgs is given by
\begin{eqnarray}
&&
{\cal M}_{even}^2=
\left(
\begin{array}{cc}
	{\cal M}_{11}^2 + \Delta_{11}
	&
	{\cal M}_{12}^2 + \Delta_{12}
	\\
	{\cal M}_{12}^2 + \Delta_{12}
	&
	{\cal M}_{22}^2 + \Delta_{22}
\end{array}
\right),
\end{eqnarray}
with
\begin{eqnarray}
{\cal M}_{11}^2 &=& M_Z^2 \cos^2  \beta \ + \ M_A^2 \sin^2 \beta, \\
{\cal M}_{12}^2 &=& - (M_A^2 + M_Z^2) \sin \beta \cos \beta, \\
{\cal M}_{22}^2 &=& M_Z^2 \sin^2 \beta \ + \ M_A^2 \cos^2 \beta,
\end{eqnarray}
where $M_A$ is the pseudo-scalar Higgs mass and $\tan \beta = v_u / v_d$.  In order to make the numerical calculations 
we use FeynHiggs~\cite{FeynHiggs} to compute the Higgs mass at two-loop level and include the one-loop corrections 
due to the existence of new leptons. These new one-loop corrections were considered in Ref.~\cite{FileviezPerez:2012iw}, 
where it has been shown that one can increase the Higgs mass in more than 5-10 GeV in a large fraction of the parameter space.
In this article we go beyond this study and show the general constraints on the supersymmetric spectra if we satisfy 
the experimental constraints on the Higgs mass.

As we have mentioned above, since we cannot predict the Higgs mass in general, we can use the recent results from ATLAS~\cite{Collaboration:2012sk,Collaboration:2012si}
and CMS~\cite{Collaboration:2012tx,Collaboration:2012tw} to constrain the allowed parameters in the theory. In order to understand
these constraints we will work in the decoupling limit in the Higgs sector, $M_A^2 \gg M_Z^2$, which has the largest contribution 
at tree level to the Higgs mass in the MSSM , and define some simple scenarios:
\begin{itemize}
\item Scenario I:  $X_{t}=X_b=0$

In this case we neglect the left-right mixing in the squark sector and take into account
only the contributions of the third generation of quark and squarks in showing the allowed 
parameter space consistent with a Higgs mass in the range $115 \ \rm{GeV} \leq M_h \leq 128$ GeV.
In order to illustrate the numerical results we scan over the ranges
\begin{center}
$200 \ \rm{GeV} \leq M_{\tilde{Q}_3},  M_{\tilde{u}_3^c}, M_{\tilde{d}_3^c} \leq 2 \ \rm{TeV}$,
\end{center}
\begin{center}
$M_{\nu_4}=M_{\nu_5}=90$ GeV and $M_{e_4}=M_{e_5}=100$ GeV,
\end{center}
\begin{center}
$0 \ \rm{GeV} \leq M_{\tilde{L}_4},  M_{\tilde{e}_4^c}, M_{\tilde{\nu}_4^c}, M_{\tilde{L}_5^c},  M_{\tilde{e}_5}, M_{\tilde{\nu}_5} \leq 1 \ \rm{TeV}$,
\end{center}
\begin{center}
$M_{Z_B}= M_{Z_L}=1 \ \rm{TeV}, \ \tan \beta_B= \tan \beta_L=2, \  L_4=- \frac{3}{2},$
\end{center}
and
\begin{center}
$100 \  \rm{GeV} \leq M_2 \leq 300 \ \rm{GeV}$, $-300 \ \rm{GeV} \leq \mu \leq 300 \ \rm{GeV}$,
$2 \leq \tan \beta \leq 6$, $M_A=1$ TeV,
\end{center}
and show the results making the calculation for the Higgs mass 
at two-loop level in the MSSM and include the new one-loop corrections in the BLMSSM. We also note that the soft mass parameters of the new sleptons can be as low as zero since they must be at least as massive as the new leptons in this case which is consistent with experimental bounds.
\item Scenario II: $X_t \neq 0$ and $X_b \neq 0$

In the second scenario we take into account the left-right mixing in the stop and sbottom sectors, 
using the same range for the input parameters as in the previous scenario and 
\begin{center}
$-4 \ \rm{TeV} \leq X_t \leq 4 \ \rm{TeV}$, $-4 \ \rm{TeV} \leq X_b \leq 4 \ \rm{TeV}$,
\end{center}

for the values of the left-right mixing in the squark sector. 
\end{itemize}

\begin{figure}[ht]
\includegraphics[scale=1,width=8.1cm]{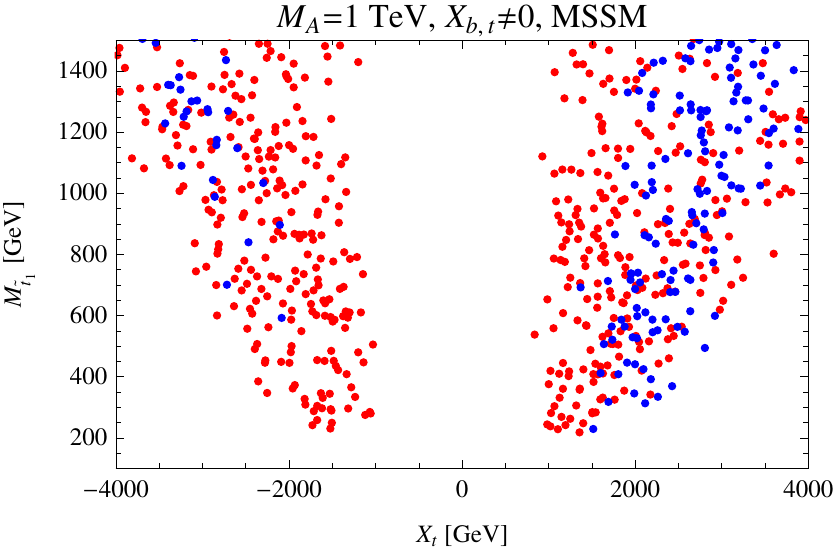}
\includegraphics[scale=1,width=8.1cm]{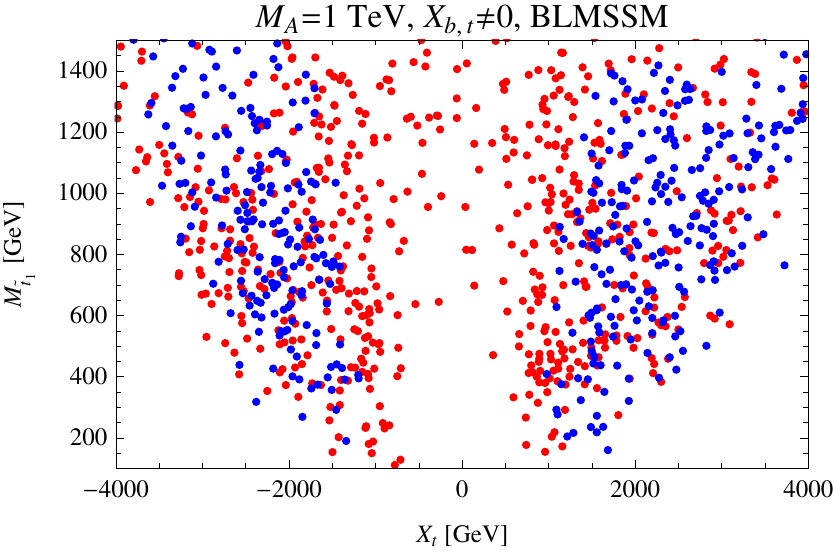}
\caption{Allowed parameter space in the MSSM and BLMSSM in the plane of lightest stop mass versus the left-right mixing in the stop sector. We use as input parameters $M_{\nu_4}=M_{\nu_5}=90$ GeV and $M_{e_4}=M_{e_5}=100$ GeV. In the MSSM we compute the Higgs mass at two-loops 
and in the BLMSSM we have the extra one-loop contributions. The red points correspond to the range when the Higgs mass is between 115 GeV and 122 GeV, while the blue points correspond to the range, 
122 GeV $\leq M_h \leq $ 128 GeV. We use $M_{\tilde{g}}=1$ TeV as the gluino mass. 
We have checked that all solutions are consistent with the bounds from the absence of color and electric charge breaking.}
\label{Fig1}
\end{figure}

In Fig.~\ref{Fig1} we show the allowed parameter space in the MSSM and BLMSSM when the Higgs mass is in the range mentioned above. 
In the MSSM we compute the Higgs mass at two-loop level using FeynHiggs and in the BLMSSM we have the extra leptonic 
one-loop contributions. Notice that the red points correspond to the range when the Higgs mass is between 115 GeV and 122 GeV, 
while the blue points correspond to the range, 122 GeV $\leq M_h \leq $ 128 GeV. We use $M_{\tilde{g}}=1$ TeV as the gluino mass.
The first main difference to notice is that in the MSSM there is no solution when $X_t=0$, while this is not the case in the BLMSSM. Therefore, for SUSY breaking scenarios such as gauge mediation where the trilinear terms are small, one can say that in the context of the BLMSSM 
it is possible to satisfy the bounds on the Higgs mass. 

\begin{figure}[ht]
\includegraphics[scale=1,width=8.3cm]{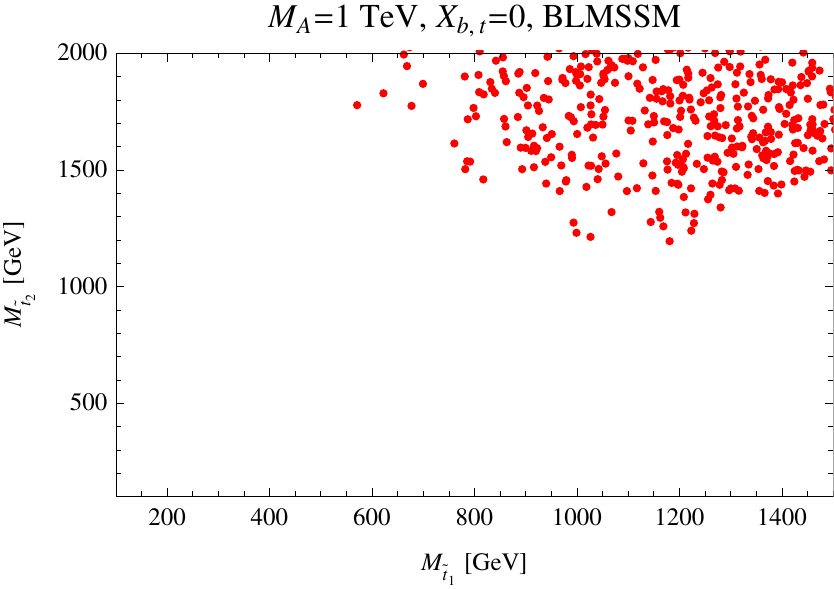}
\includegraphics[scale=1,width=8.0cm]{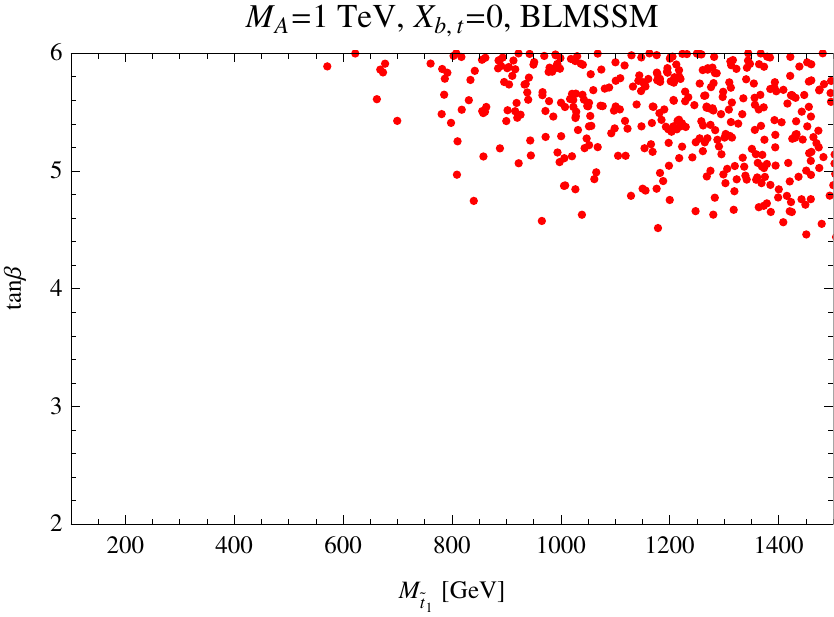}
\caption{Allowed parameter space in the BLMSSM in zero mixing scenario 
$X_t=X_b=0$. We use as input parameters $M_{\nu_4}=M_{\nu_5}=90$ GeV and $M_{e_4}=M_{e_5}=100$ GeV.}
\label{Fig2}
\end{figure}

In Fig.~\ref{Fig2} we show the allowed parameter space in scenario I for the BLMSSM. Notice that in this case the lightest stop 
can be as light as 600 GeV, while the heaviest stop is always above 1 TeV. For our input parameters we find allowed 
solutions when $\tan \beta$ is larger than 4, since the tree level mass is directly related to $\tan \beta$.
However, as we have mentioned before, there is an upper bound on $\tan \beta$ coming from perturbativity, and combining these 
two bounds limits allowed range to a small region.

\begin{figure}[ht]
\includegraphics[scale=1,width=8.0cm]{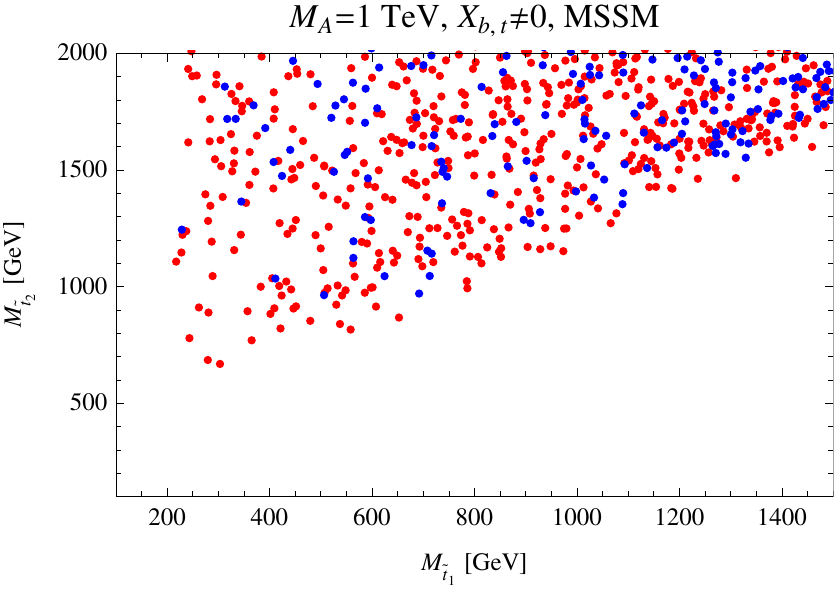}
\includegraphics[scale=1,width=8.0cm]{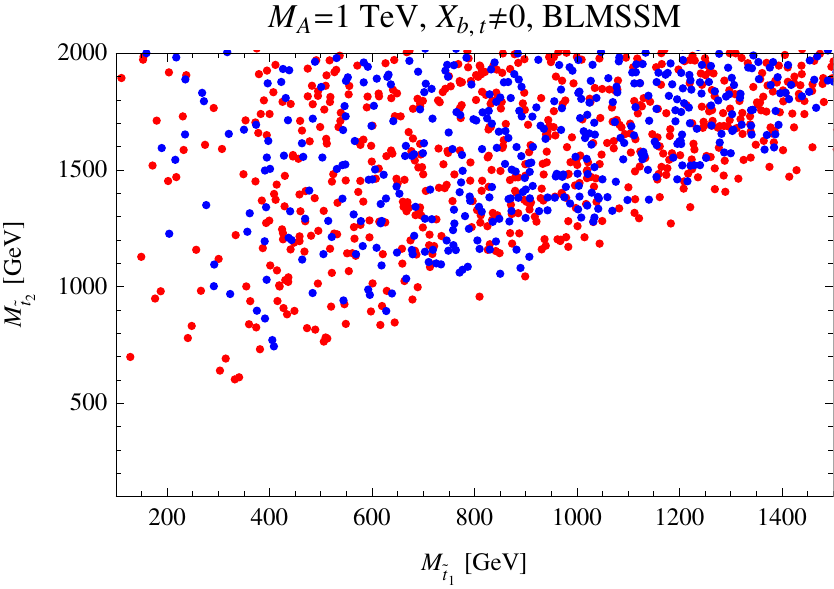}
\caption{Allowed parameter space in the MSSM and BLMSSM in the non-zero mixing scenario 
$X_t\neq 0$ and $X_b \neq 0$. We use as input parameters $M_{\nu_4}=M_{\nu_5}=90$ GeV and $M_{e_4}=M_{e_5}=100$ GeV.}
\label{Fig3}
\end{figure}

\begin{figure}[ht]
\includegraphics[scale=1,width=8.0cm]{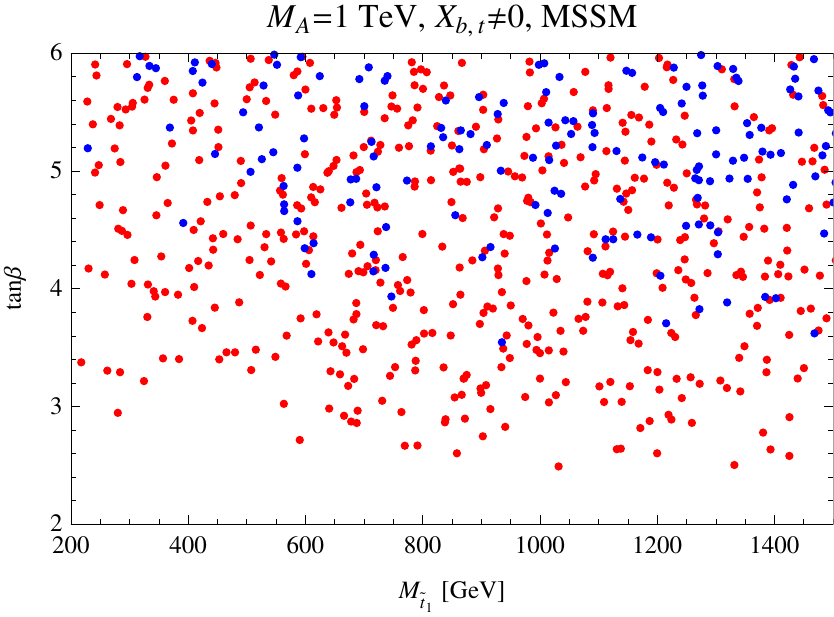}
\includegraphics[scale=1,width=8.0cm]{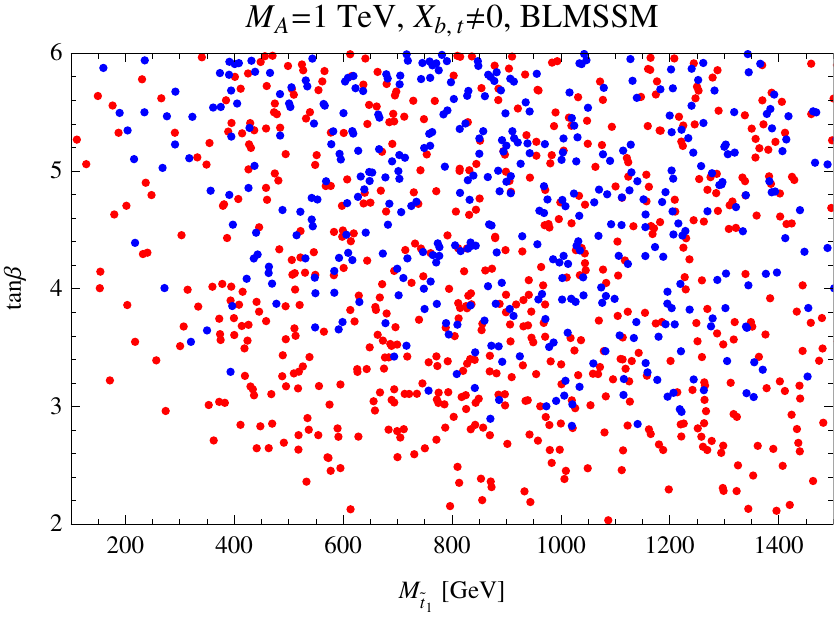}
\caption{Allowed parameter space in the MSSM and BLMSSM in the non-zero mixing scenario 
$X_t\neq 0$ and $X_b \neq 0$. We use as input parameters $M_{\nu_4}=M_{\nu_5}=90$ GeV and $M_{e_4}=M_{e_5}=100$ GeV.}
\label{Fig4}
\end{figure}

In scenario II the left-right mixing in the squark sector can be large and we see in Fig.~\ref{Fig3} that the lightest stop can 
be very light, in the 100 GeV region, consistent with the Higgs bounds. The situation is similar in both models, in the 
MSSM and the BLMSSM. However, in the BLMSSM we find more solutions which correspond to the Higgs mass in 
the range, 122 GeV $\leq M_h \leq $ 128 GeV, due to the contributions of the new leptons. Notice, that the heaviest stop 
can be as light as 500 GeV in both models. In this scenario we do not find any relevant lower bound on $\tan \beta$ 
since it is easier to satisfy the Higgs bounds.  See Fig.~\ref{Fig4} for the numerical results in the $\tan \beta$-$M_{\tilde{t}_1}$ plane. 
\section{Radiative Higgs Decays: $h \to \gamma \gamma$ and $h \to gg$}
As it is well-known, the excess reported by CMS and ATLAS are in the $\gamma \gamma$ 
channel, while the other channels are mainly SM-like. Now, knowing the allowed parameter 
space consistent with a Higgs mass, $M_h = 115 - 128$ GeV, we can 
show the predictions for the radiative Higgs decays.

\begin{figure}[h]
\includegraphics[scale=1,width=8.1cm]{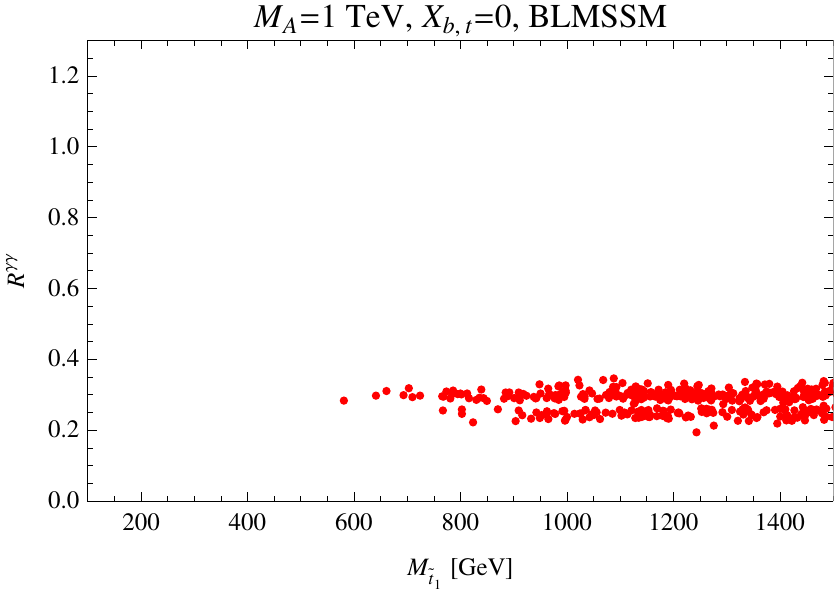}
\caption{Predictions for $R^{\gamma \gamma}$ in scenario I.}
\label{Fig5}
\end{figure}

The Higgs decays at tree level are not modified, but the radiative Higgs decays, 
$h \to \gamma \gamma$, $h \to gg$ and $h \to Z \gamma$ can be modified due to the existence 
of new leptons and their superpartners. For the study of the Higgs decay into gamma gamma
 in the decoupling limit in the MSSM see Ref.~\cite{Djouadi:1996pb}. 
In Fig.~\ref{Fig5} we show the predictions for the ratio $R^{\gamma \gamma}$ defined as
\begin{equation}
R^{\gamma \gamma}= \frac{\Gamma(h \to \gamma \gamma)}{\Gamma(h \to \gamma \gamma)_{SM}},
\end{equation}
where we have scanned over the ranges mentioned in the previous section and assume 
zero left-right mixing. Here we show only the predictions in the BLMSSM because in the MSSM 
one cannot satisfy the Higgs bounds. Notice that the ratio $R^{\gamma \gamma}$ is around 0.3 
due to the suppression of the new leptons and their superpartners. In the non-supersymmetric 
version of the model this effect was studied in Ref.~\cite{Ishiwata:2011hr}. In our case the new sleptons further
suppress this ratio if they are very light.

\begin{figure}[h]
\includegraphics[scale=1,width=8.1cm]{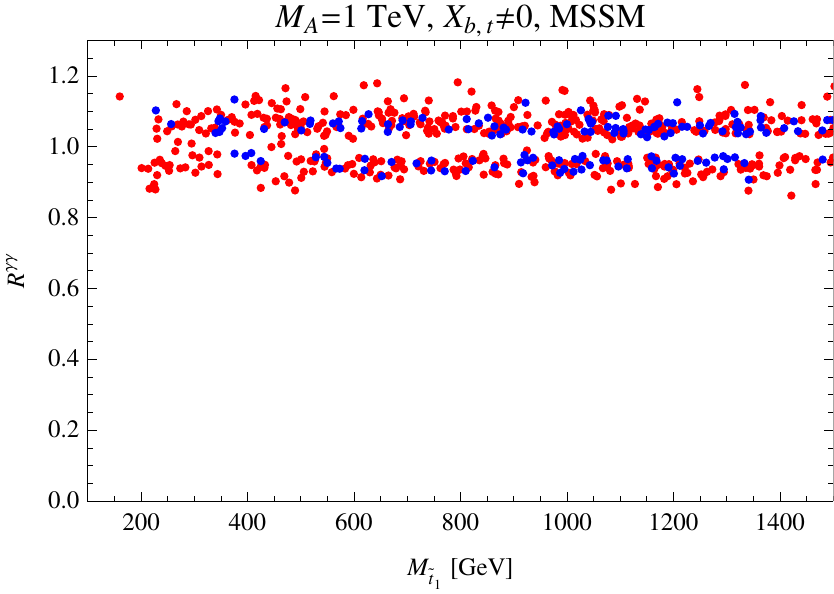}
\includegraphics[scale=1,width=8.1cm]{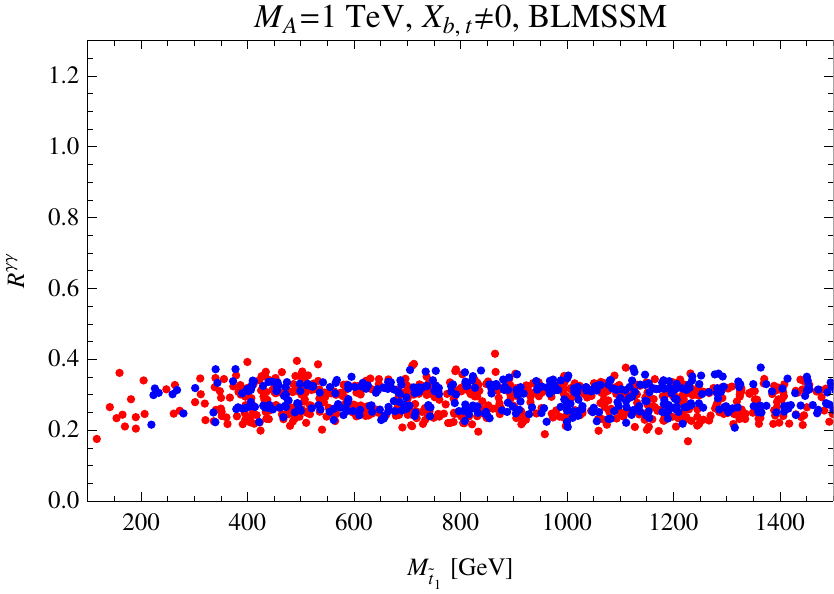}
\caption{Predictions for $R^{\gamma \gamma}$ when $X_t \neq 0$ and $X_b \neq 0$.}
\label{Fig6}
\end{figure}

\begin{figure}[h]
\includegraphics[scale=1,width=8.1cm]{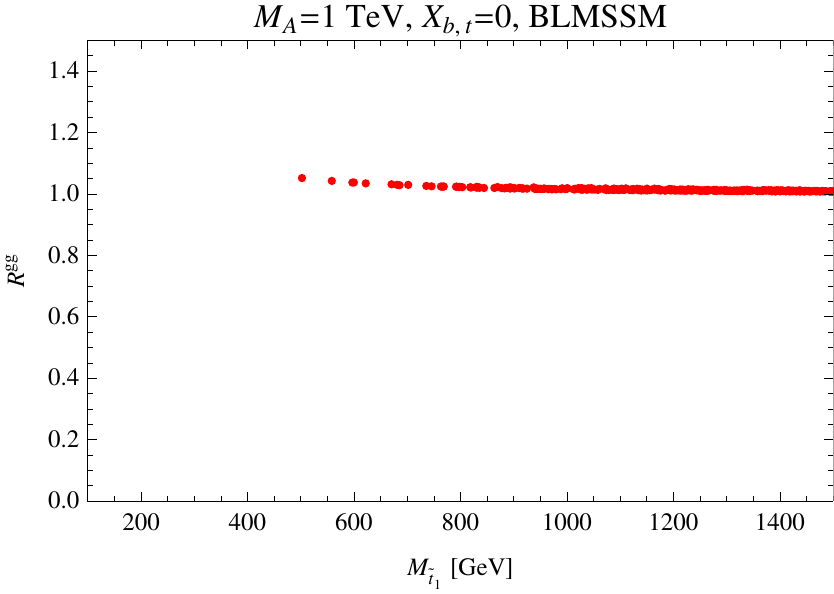}
\caption{Predictions for $R^{gg}$ in scenario I.}
\label{Fig7}
\end{figure}

\begin{figure}[h]
\includegraphics[scale=1,width=8.1cm]{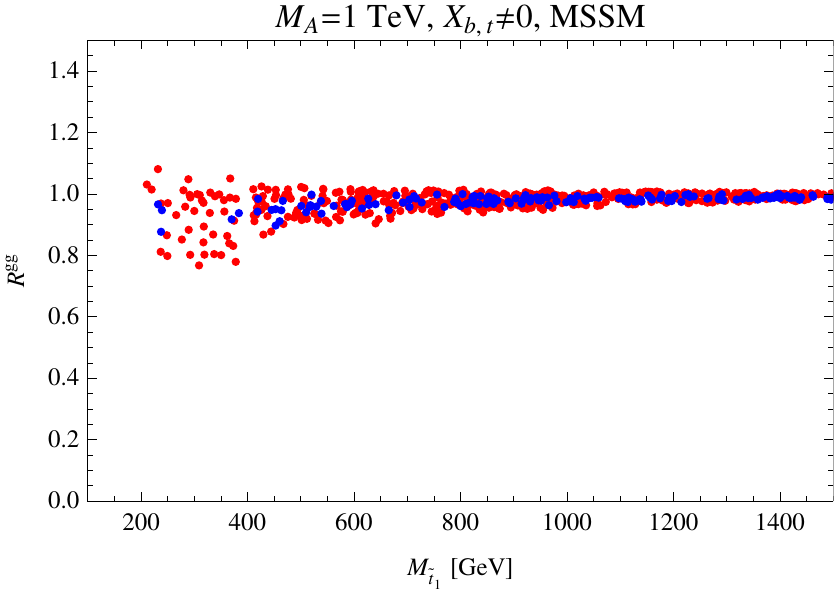}
\includegraphics[scale=1,width=8.1cm]{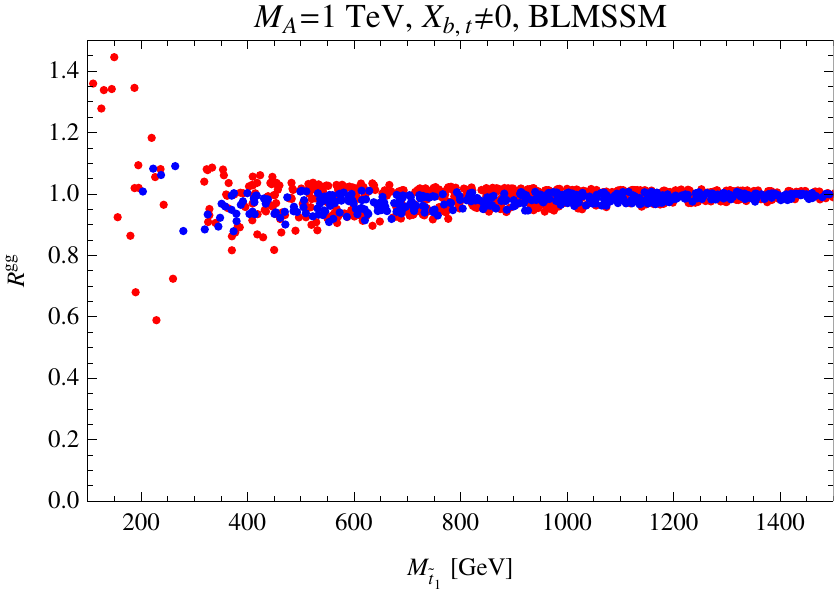}
\caption{Predictions for $R^{gg}$ when $X_t \neq 0$ and $X_b \neq 0$.}
\label{Fig8}
\end{figure}

In scenario II the situation is quite different because the stops can be very light and the left-right mixing 
can play a role is the enhancement of the $R^{\gamma \gamma}$ ratio. In Fig.~\ref{Fig6} we see that in the MSSM 
the predictions are SM-like but the ratio can change between 0.85 and 1.2 in the whole parameter space. 
In the BLMSSM the situation is different since $R^{\gamma \gamma}$ ratio can be between 0.1 and 0.4.
Therefore, one can say that in the BLMMSM one expects a large suppression for the 
$gg \to h \to \gamma \gamma$ signals. In our opinion, since still the experimental collaborations do not have 
enough results to claim a discovery, we only take these results as a hint against this model. 
In order to understand the impact of the SUSY spectrum on the Higgs signals we need to study the radiative decays, $h \to gg$. 
We define the quantity
\begin{equation}
R^{gg}= \frac{\Gamma(h \to g g)}{\Gamma(h \to g g)_{SM}},
\end{equation} 
and show the predictions in scenario I and II in Figs.~\ref{Fig7} and~\ref{Fig8}, respectively. It is easy to understand 
that in scenario I this ratio is not modified because the stops are heavy. In the second scenario the situation 
is different because the stops can be light and the left-right mixing can change the sign of the stop contribution.
These results are shown in Fig.~\ref{Fig8}, where we can see that in the MSSM the ratio $R^{gg}$ can be between 0.6 and 1, 
while in the BLMSSM the range can be 0.2-1.3 when the stop is very light and the left-right mixing is large. 
It is important to mention that the $R^{gg}$ cannot be too different from the SM because at the moment there 
are not large excesses in other channels where the Higgs decays into two gauge bosons.  

Since the relevant quantity for the experiments is defined as
\begin{equation}
C_{\gamma \gamma}=\frac{\sigma (gg \to h) \times \rm{Br}(h \to \gamma \gamma)}{\sigma (gg \to h)_{SM} \times \rm{Br}(h \to \gamma \gamma)_{SM}} \approx \frac{\Gamma (h \to gg ) \times \rm{Br}(h \to \gamma \gamma)}{\Gamma (h \to gg )_{SM} \times \rm{Br}(h \to \gamma \gamma)_{SM}},
\end{equation}
where we've used the narrow-width approximation where the cross section is proportional to the Higgs decay width into two gluons, see for example~\cite{Spira:1997dg},
we will use the last part of the above expression to calculate the predictions made by the MSSM and the BLMSSM. 

\begin{figure}[h]
\includegraphics[scale=1,width=8.1cm]{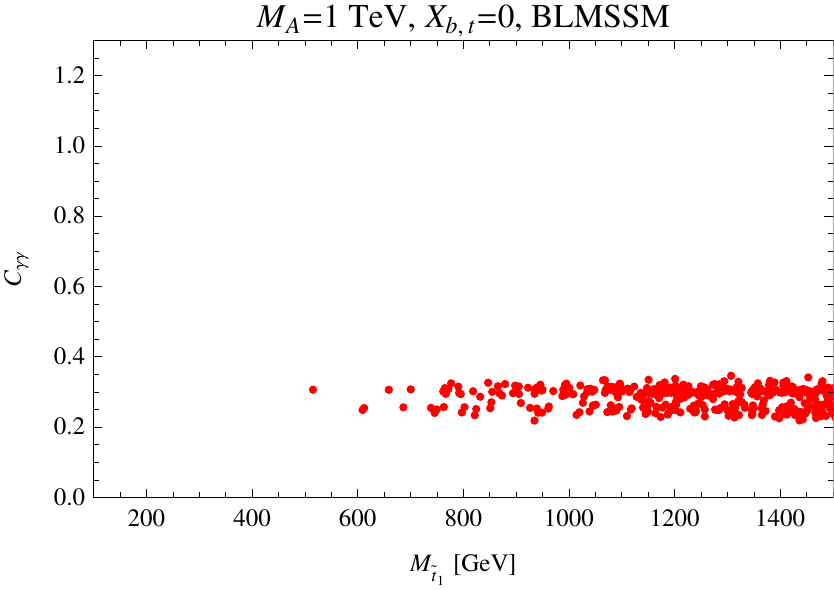}
\caption{Predictions for $C_{\gamma \gamma}$ in scenario I.}
\label{Fig9}
\end{figure}

\begin{figure}[h]
\includegraphics[scale=1,width=8.1cm]{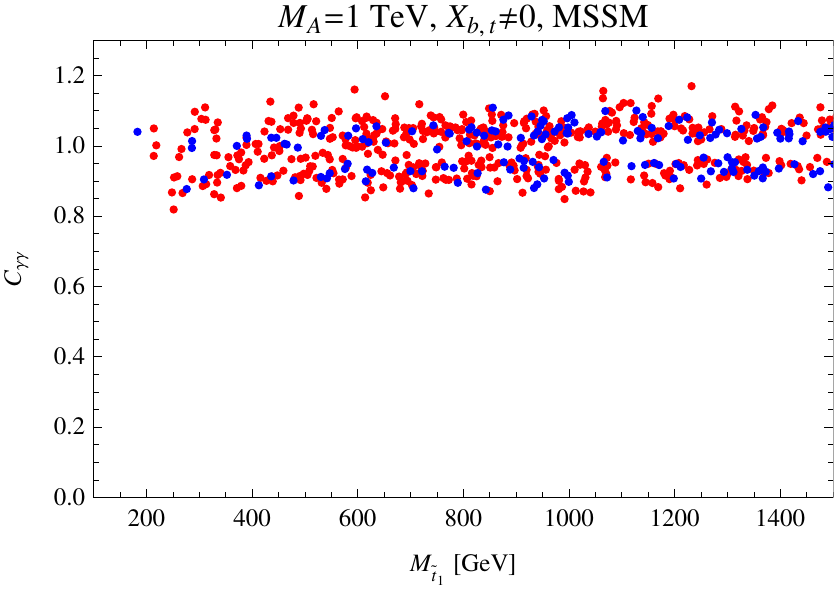}
\includegraphics[scale=1,width=8.1cm]{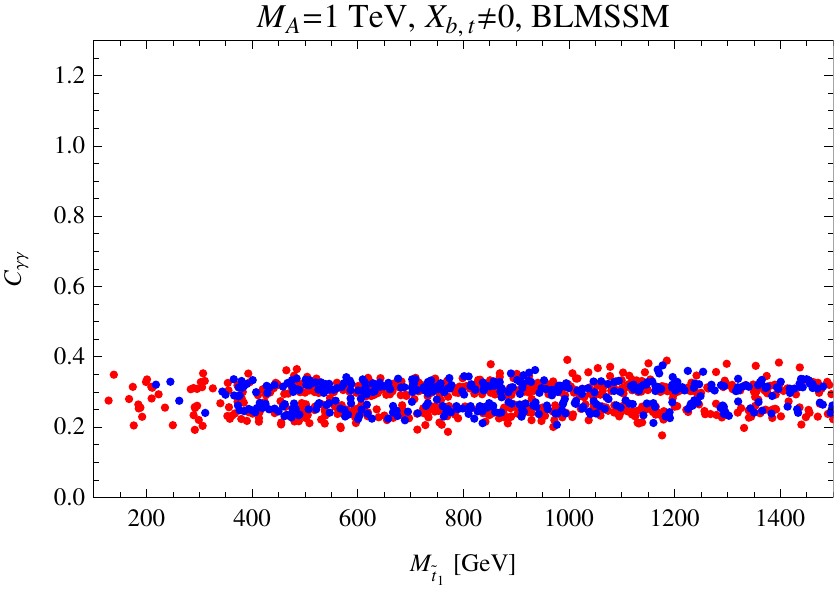}
\caption{Predictions for $C_{\gamma \gamma}$ when $X_t \neq 0$ and $X_b \neq 0$.}
\label{Fig10}
\end{figure}

Knowing the results in Figs.~\ref{Fig5}-\ref{Fig8}, we can show the predictions for the $C_{\gamma \gamma}$.
Since $\gamma \gamma$ and $gg$ ratios are SM-like in the MSSM in Fig.\ref{Fig10} we see that $C_{\gamma \gamma}$ is SM-like, but it can change between 
0.8 and 1.1. In the BLMSSM we just extrapolate the suppression in the $\gamma \gamma$ channel to see that there is a suppression for the signals in this channel.
See Figs.~\ref{Fig9} and~\ref{Fig10} for details.

In summary, we can see that in the context of the BLMMSM one predicts less events in the $gg \to h \to \gamma \gamma$ channel. 
Therefore, one could rule out this model in the near future if the signals around, $M_h \sim 125$ GeV, are confirmed by the LHC experiments.
\section{Evolution of the Gauge and Yukawa Couplings}
\begin{flushleft}
The evolution of the gauge couplings at one-loop level is given by the well-known expression
\end{flushleft}
\begin{equation}
	\frac{1}{\alpha_a(\Lambda_1)} = \frac{1}{\alpha_a(\Lambda_2)} \ + \ \frac{b_a}{2 \pi} \  \rm{Log}  \left( \frac{\Lambda_2}{\Lambda_1} \right),
\end{equation}
where $\alpha_a=g_a^2/4 \pi$, $b_a$ are the beta functions for the different groups and $\Lambda_i$ is a given scale.  In order to make the numerical study we will assume that only the new 
leptons exist below the SUSY scale while the thresholds associated with the new sleptons, new quarks and new squarks are numerically close to the SUSY scale. 

The new leptons therefore effect the $b_a$ values of the SM gauge group they are:
\begin{equation}
	b_3 = -7, \quad b_2 = -\frac{15}{6}, \quad b_1 = \frac{53}{10}, 
\end{equation}
while above the SUSY scale:
\begin{equation}
	b_3 = 1, \quad b_2 = 5, \quad b_1 = \frac{53}{5},
\end{equation}
\begin{equation}
	b_B = N_B \left(26 B_4^2 + \frac{80}{3} B_4 + \frac{170}{9} \right), \quad b_L = N_L \left(8 L_4^2 + 24 L_4 +56 \right).
\end{equation}
\begin{figure}[ht!]
	\includegraphics[scale=1,width=10.5cm]{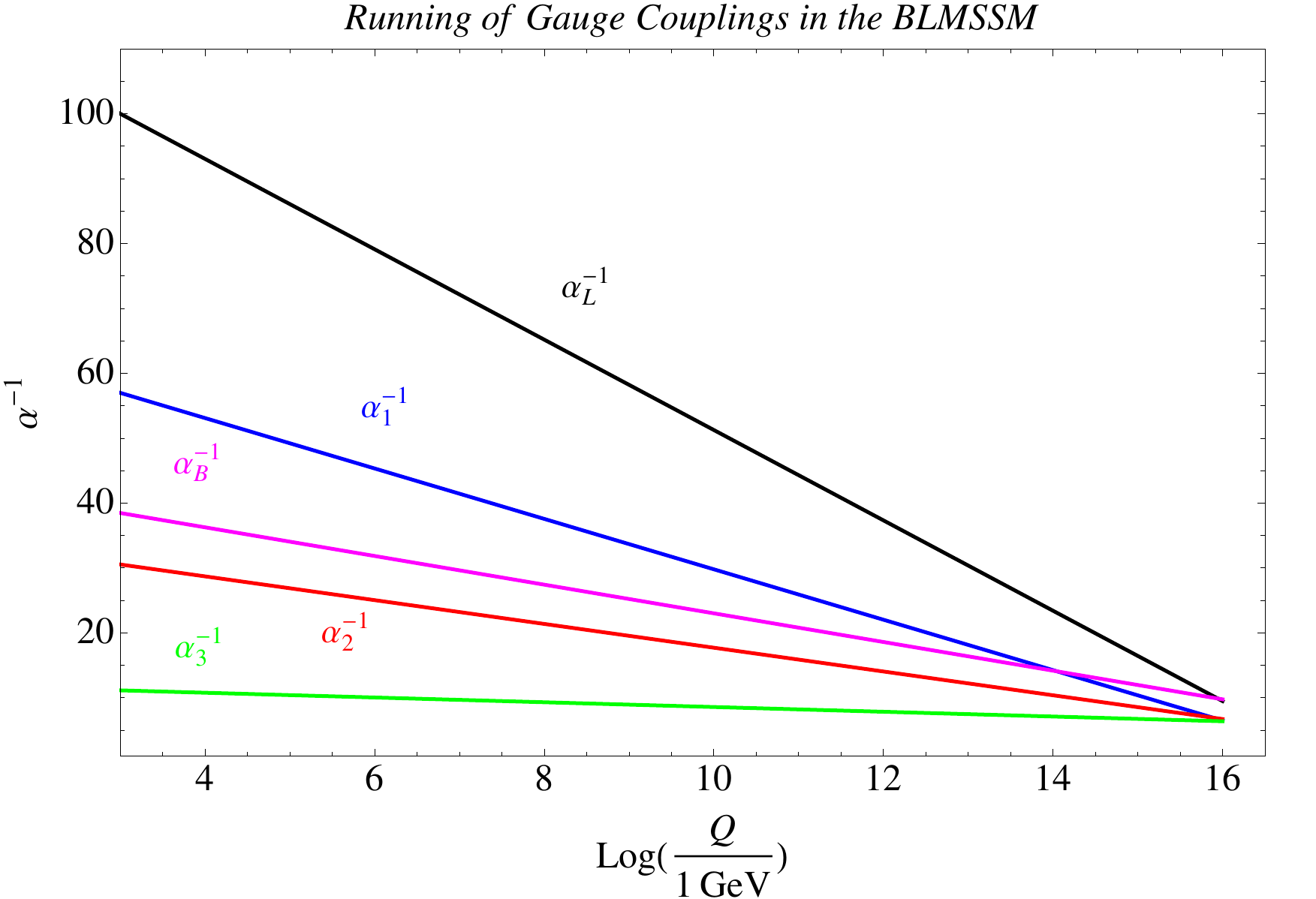}
	\caption
	{Running of the gauge couplings in the BLMSSM assuming $\alpha_B=0.026$ and $\alpha_L=0.01$ at the SUSY scale. }	
	\label{Fig11}
\end{figure}
Where $N_B$ and $N_L$ are the normalizations for $U(1)_B$ and $U(1)_L$. While its unclear what these values should be without knowledge of the high scale physics, we will use, for simplicity, $N_B = N_L = \frac{1}{2}$. Also using $B_4=-1/2$ and $L_4=-3/2$ we show in Fig.~\ref{Fig11} the running of the gauge couplings when $\alpha_B=0.026$ 
and $\alpha_L=0.01$ at the SUSY scale. As one can see from these results we can keep the unification of the gauge 
couplings of the MSSM and we can have a simple solution for the unification of the $\alpha_L$ and $\alpha_B$ 
at the scale $10^{16}$ GeV. In this way one can imagine a possible unified theory at the high scale.
This type of GUT model will be investigated in a future publication.

\begin{figure}[ht!]
	\includegraphics[scale=1,width=7.5cm]{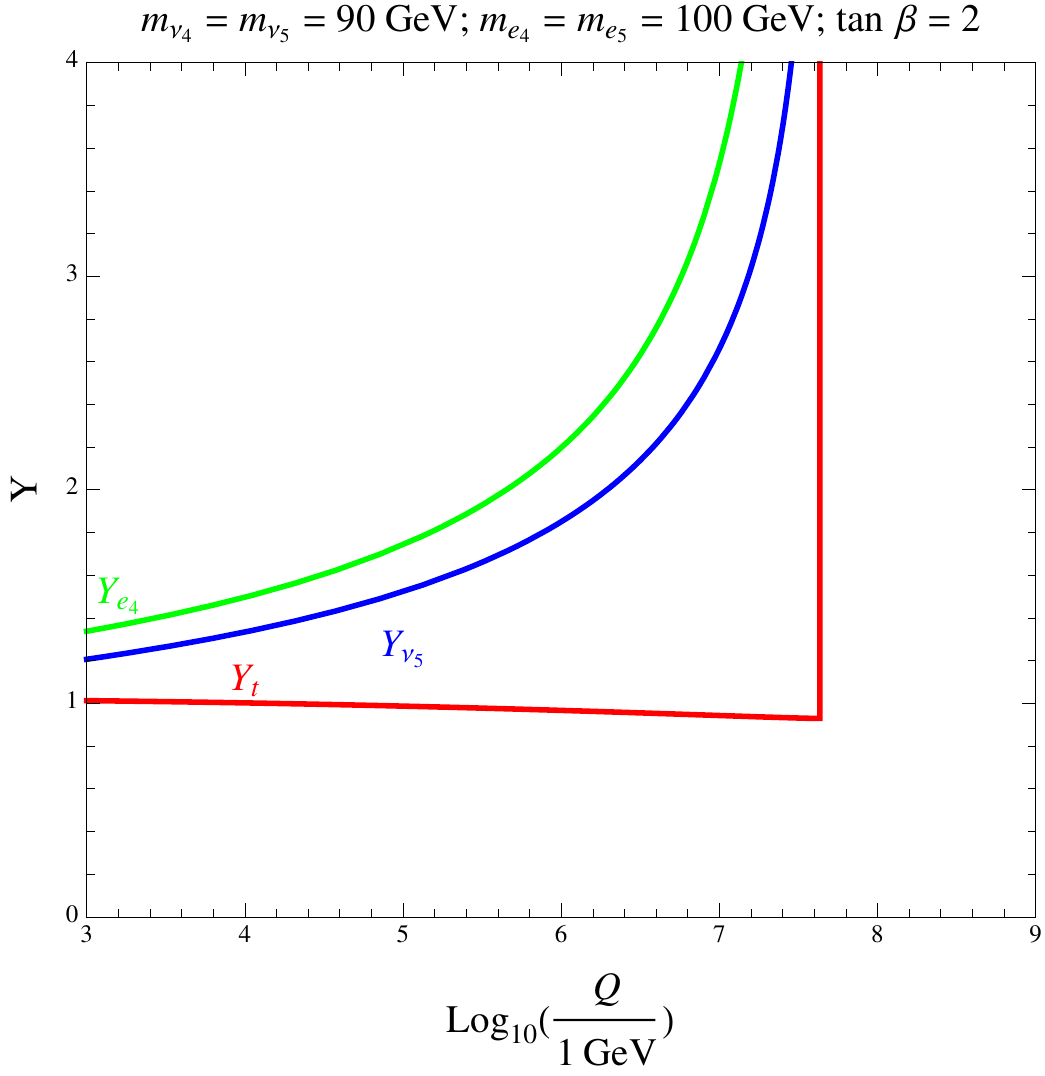}
	\includegraphics[scale=1,width=7.5cm]{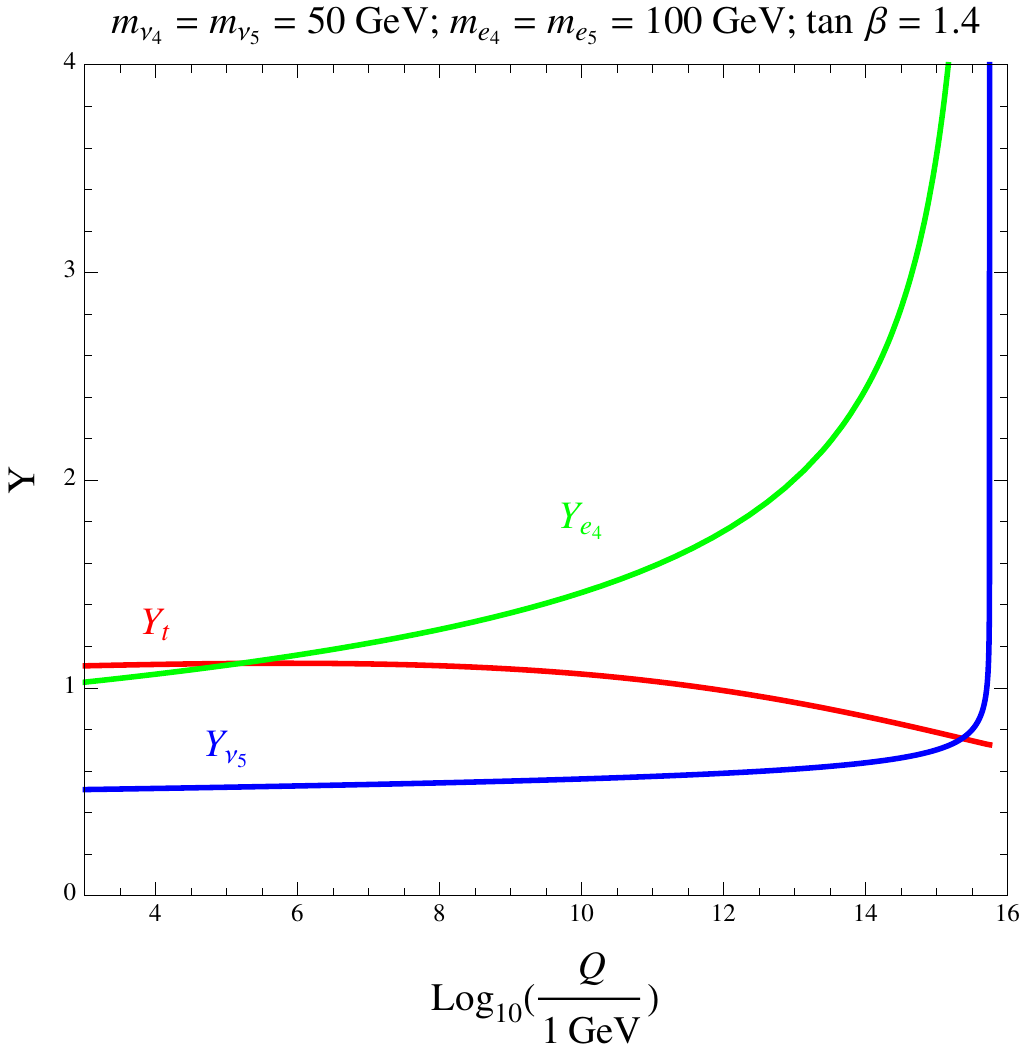}
	\caption
	{Yukawa coupling running in the BLMSSM for $Y_t, Y_{e_4}$ and $Y_{\nu_5}$ above the SUSY scale 
	for $m_{\nu_4} = m_{\nu_5} = 90 (50)$ GeV and $m_{e_4} = m_{e_5} = 100$ GeV and for $\tan \beta = 2 (1.4)$ 
	with gauge couplings as in Fig.~\ref{alpha}.}
	\label{Fig12}
\end{figure}

\begin{figure}[h]
	\includegraphics[scale=1,width=7.7cm]{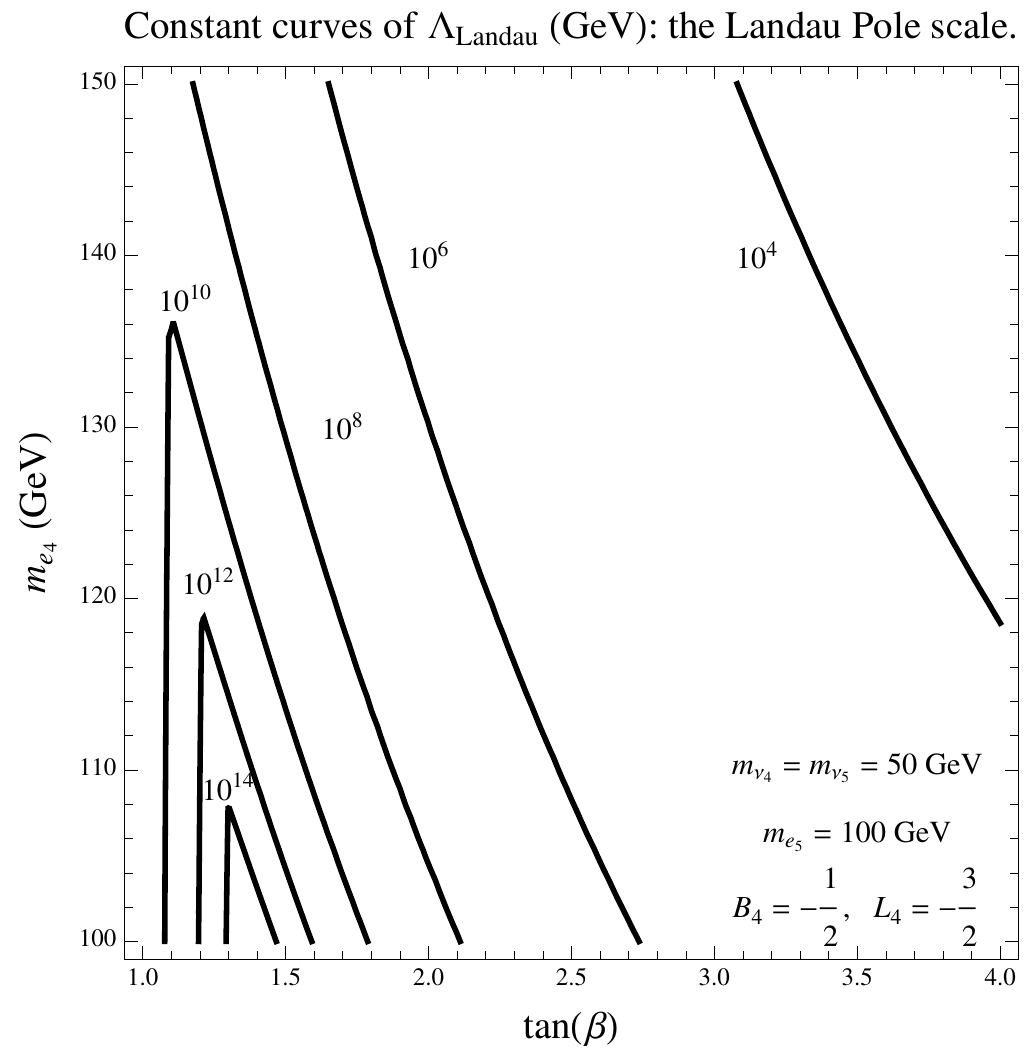}
	\includegraphics[scale=1,width=7.6cm]{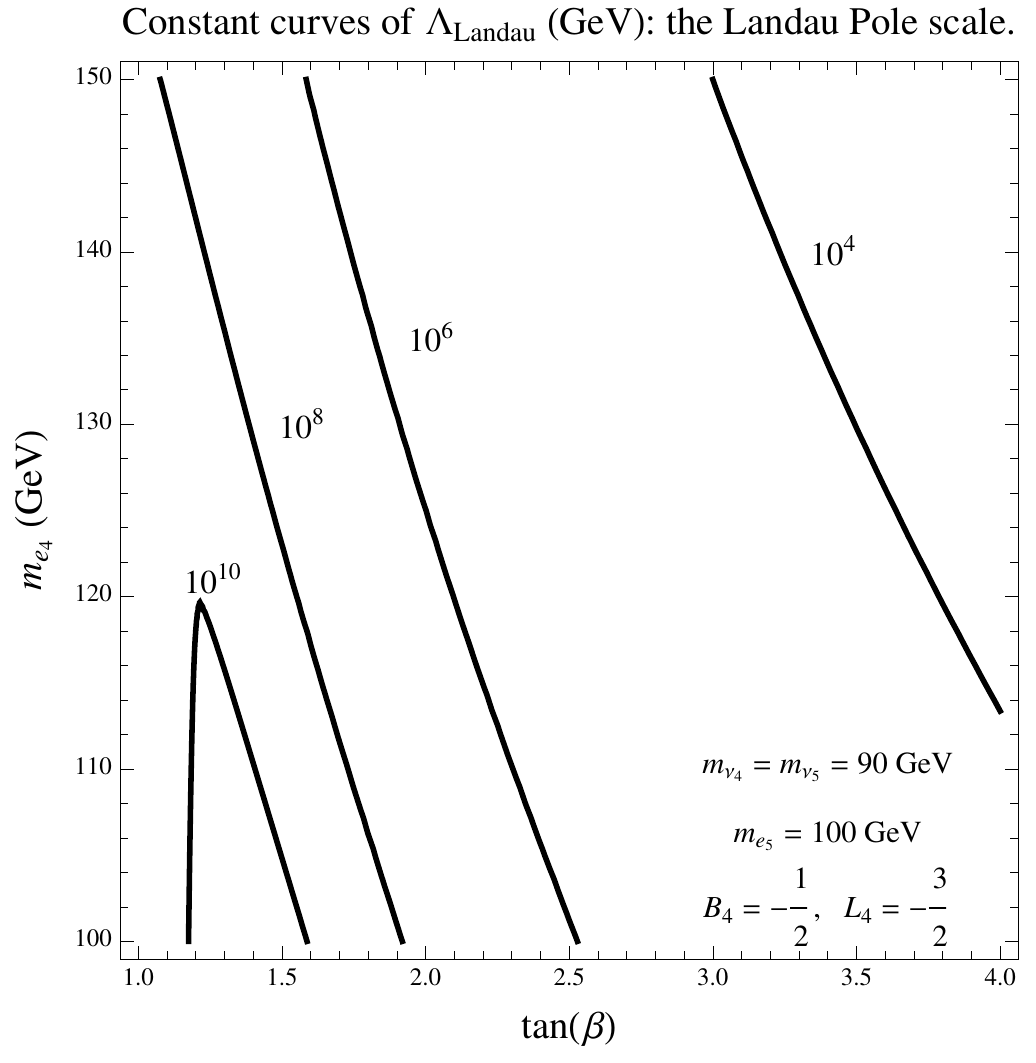}
	\caption
	{Isoplot for the scale where one has a Landau pole in the $m_{e_4}$-$\tan \beta$ plane.}
	\label{Fig13}
\end{figure}

In order to study the possible existence of a Landau pole we study here the evolution of the Yukawa couplings.
See the Appendix for the renormalization group of equations for these couplings. The matching conditions 
for the Yukawa couplings at the SUSY scale are
\begin{equation}
	Y_i = \frac{h_i}{\sin \beta}, \quad  Y_{j} = \frac{h_j}{\cos \beta}, \quad i=t, \nu_4, e_5, \quad  j=b, \tau, e_4, \nu_5,
\end{equation}
therefore giving a boost to the latter set of Yukawa couplings for $\tan \beta > 1$. In Fig.~\ref{Fig12} we show the evolution 
of the largest Yukawa couplings ($Y_t, Y_{e_4}, Y_{\nu_5}$) for two different scenarios: a) $m_{\nu_4} = m_{\nu_5} = 90$ GeV and $\tan \beta = 2$,
b) $m_{\nu_4} = m_{\nu_5} = 50$ GeV, $m_{e_4} = m_{e_5} = 100$ GeV and for $\tan \beta =1.4$. In the first 
scenario the Landau pole is around $10^{7}$ GeV, while when $\tan \beta = 1.4$ there is a Landau pole at the scale, 
$10^{14}$ GeV. In order to show more general results we show in Fig.~\ref{Fig13} the isoplot for the scale where we have 
a Landau pole in the $m_{e_4}$-$\tan \beta$ plane. It is important to mention that in general the cutoff of theory 
can be very large. However, since the local baryon number is broken at the low scale we do not need to assume 
a large cutoff or the desert in order to satisfy the bounds on the proton decay lifetime. 
%
\section{Summary}
We have discussed the main features of a simple extension of the minimal supersymmetric standard model 
where the baryon and lepton numbers are local symmetries. We refer to this theory as the ``BLMSSM". 
In this context we do not need to assume a large desert between the electroweak scale and grand unified scale 
in order to satisfy the proton decay bounds even if B and L are broken at the low scale.
In this context the lepton number is broken in an even number while the baryon number violating operators 
can change B in one unit. There is no flavour violation at tree level due to absence of mixing between the 
SM fermion and new families and Landau pole at the low scale. The light Higgs boson mass 
can be large without assuming a large stop mass and left-right mixing and one could modify the current 
LHC bounds on the supersymmetric spectrum due to the presence of the baryon number violating interactions.

In section III we discussed the constraints on the $\lambda_{ijk}^{''}$ from cosmology and the impact of these couplings 
on the LHC searches for supersymmetric particles. In this case we can have very interesting signals without missing energy.
For example, if the stop is the lightest supersymmetric particle one can have signals with displaced vertices and four jets.
These are interesting signals can shed light on the possibility to have light stops in the spectra.

We have investigated in great detail the correlation between the light Higgs mass and the decay of the Higgs boson into 
two photons following the new results presented by the ATLAS and CMS collaborations. In this theory the new light leptons 
modify appreciably the predictions of the Higgs decays into two gammas. The contraints on the Higgs mass tell us how light 
the lightest stop, $M_{\tilde{t}_1}$, can be and since B is broken we can satisfy the collider bounds.  
We have found that in the context of the BLMMSM one predicts less events in the $gg \to h \to \gamma \gamma$ channel. 
Therefore, one could rule out this model in the near future if the signals around, $M_h \sim 125$ GeV, are confirmed by the LHC experiments.

In this theory the fields $X$ and $\bar{X}$ (or their superpartners) can be a cold dark matter candidate if there are the lightest fields with baryon number 
and do not get a vacuum expectation value. This is true even when R-parity is broken in this theory. This is an interesting 
result which we will be investigated in a future publication. See Ref.~\cite{Dulaney:2010dj} for the study of this dark matter 
candidate in a non-supersymmetric version of the model.

The running of the Yukawa couplings were studied in order to understand the possible cutoff of the theory. We have found 
that for small values of $\tan \beta$ the cutoff of the theory can be very large. It is important to mention that since the baryon 
and lepton numbers are broken at the low scale there is no need to have a large cutoff. In summary, we could say that the 
BLMSSM is a consistent theory where one could expect a light stop-sbottom spectrum in agreement with all experiments 
and predicts new light leptons. The collider signals of this theory will be investigated in a future publication.

\vspace{0.5cm}
{\textit{Acknowledgments}}:
We would like to thank Mark B. Wise for discussions and many interesting comments. 
The work of P. F.P. has been supported by the James Arthur Fellowship, CCPP, New York University. 
The work of JMA and BF was supported in part by the U.S. Department of Energy under contract No. DE-FG02-92ER40701.
\appendix
\section{Particle Content}
\begin{table}[htdp]
\begin{center}
\begin{tabular}{|c|c|c|c|c|c|}
\hline
Superfields & $SU(3)_C$ & $SU(2)_L$ & $U(1)_Y$ & $U(1)_B$ & $U(1)_L$\\
\hline
\hline
 $\hat{Q}_i$ & 3 & 2 & 1/6 & 1/3 & 0 \\
 \hline
 $\hat{u}^c_i$ & $\bar{3}$ & 1 & -2/3 & -1/3 & 0 \\
 \hline
$\hat{d}^c_i$ & $\bar{3}$ & 1 & 1/3 & -1/3 & 0 \\
\hline
$\hat{L}_i$ & 1 & 2 & -1/2 & 0 & 1\\
\hline
$\hat{e}^c_i$ & 1 & 1 & 1 & 0 & -1 \\
\hline
$\hat{\nu}^c_i$ & 1 & 1 & 0 & 0 & -1 \\
\hline
$\hat{Q}_4$ & 3 & 2 & 1/6 & $B_4$ & 0 \\
\hline
$\hat{u}^c_4$ & $\bar{3}$ & 1 & -2/3 & $- B_4$ & 0 \\
\hline
$\hat{d}^c_4$ & $\bar{3}$ & 1 & 1/3 & $- B_4$ & 0 \\
\hline
$\hat{L}_4$ & 1 & 2 & -1/2 & 0 & $L_4$ \\
\hline
$\hat{e}^c_4$ & 1 & 1 & 1 & 0 & -$L_4$ \\
\hline
$\hat{\nu}^c_4$ & 1 & 1 & 0 & 0 & -$L_4$ \\
\hline
$\hat{Q}_5^c$ & $\bar{3}$ & 2 & -1/6 & $-1 - B_4$ & 0 \\
\hline
$\hat{u}_5$ & $3$ & 1 & 2/3 &  $1 + B_4$ & 0 \\
\hline
$\hat{d}_5$ & $3$ & 1 & -1/3 & $1 + B_4$ & 0 \\
\hline
$\hat{L}_5^c$ & 1 & 2 & 1/2 & 0 & $-3 - L_4$ \\
\hline
$\hat{e}_5$ & 1 & 1 & -1 & 0 & $3 + L_4$ \\
\hline
$\hat{\nu}_5$ & 1 & 1 & 0 & 0 & $3 + L_4$ \\
\hline
$\hat{H}_u$ & 1 & 2 & 1/2 & 0 & 0 \\
\hline
$\hat{H}_d$ & 1 & 2 & -1/2 & 0 & 0 \\
\hline
$\hat{S}_B$ & 1 & 1 & 0 & 1 & 0 \\
\hline
$\hat{\bar{S}}_B$ & 1 & 1 & 0 & -1 & 0 \\
\hline
$\hat{S}_L$ & 1 & 1 & 0 & 0 & -2 \\
\hline
$\hat{\bar{S}}_L$ & 1 & 1 & 0 & 0 & 2 \\
\hline
$\hat{X}$ & 1 & 1 & 0 & $2/3 + B_4$ & 0 \\
\hline
$\hat{\bar{X}}$ & 1 & 1 & 0 & $-2/3 - B_4$ & 0 \\
\hline
\end{tabular}
\end{center}
\label{default}
\caption{Superfields in the BLMSSM. The index $i=1,2,3$.}
\end{table}%
\section{Sfermion Masses}
The sbottom mass matrix is defined by
\begin{eqnarray}
&&
{\cal M}_{\tilde{b}}^2=
\left(
\begin{array}{cc}
      M_b^2 + M_{\tilde{Q}_3}^2 \ - \ \left( \frac{1}{2} - \frac{1}{3}  \sin^2 \theta_W \right) M_Z^2 \cos 2 \beta +  \frac{1}{3} D_B
	&
	M_b  \ X_b
	\\
	M_b \ X_b
	&
	M_b^2 \ + \ M_{\tilde{d}_3^c}^2 - \frac{1}{3} \sin^2 \theta_W M_Z^2 \cos 2 \beta  -  \frac{1}{3} D_B
\end{array}
\right), \nonumber \\
\end{eqnarray}
where $M_{\tilde{d}_3^c}^2$ is a soft mass, $D_B=\frac{1}{2} M_{Z_B}^2 \cos 2 \beta_B$ and $X_b = A_b - \mu \tan \beta$. The mass matrix for the new sleptons is given by
\begin{eqnarray}
&&
{\cal M}_{\tilde{e}_4}^2=
\left(
\begin{array}{cc}
      M_{e_4}^2 + M_{\tilde{L}_4}^2 \ - \ \left( \frac{1}{2} -  \sin^2 \theta_W \right) M_Z^2 \cos 2 \beta +  L_4 D_L
	&
	M_{e_4}  \ X_{e_4}
	\\
	M_{e_4} \ X_{e_4}
	&
	M_{e_4}^2 \ + \ M_{\tilde{e}_4^c}^2 - \sin^2 \theta_W M_Z^2 \cos 2 \beta  -  L_4 D_L
\end{array}
\right), \nonumber \\
\end{eqnarray}
where $M_{\tilde{L}_4}^2$ and $M_{\tilde{e}_4^c}^2$ are soft masses, $D_L=-\frac{1}{4} M_{Z_L}^2 \cos 2 \beta_L$
and $X_{e_4}=A_{e_4} - \mu \tan \beta$.
The mass matrix for the fourth generation heavy neutrino is given by
\begin{eqnarray}
&&
{\cal M}_{\tilde{\nu}_4}^2=
\left(
\begin{array}{cc}
      M_{\nu_4}^2 + M_{\tilde{L}_4}^2 \ + \  \frac{1}{2} M_Z^2 \cos 2 \beta +  L_4 D_L
	&
	M_{\nu_4}  \ X_{\nu_4}
	\\
	M_{\nu_4} \ X_{\nu_4}
	&
	M_{\nu_4}^2 \ + \ M_{\tilde{\nu}_4^c}^2 -  L_4 D_L
\end{array}
\right), \nonumber \\
\end{eqnarray}
where $X_{\nu_4}=A_{\nu_4} - \mu \cot \beta$.
In the case of the leptons of the fifth generation the mass matrices read as
\begin{eqnarray}
&&
{\cal M}_{\tilde{e}_5}^2=
\left(
\begin{array}{cc}
      M_{e_5}^2 + M_{\tilde{L}_5^c}^2 \ + \ \left( \frac{1}{2} -  \sin^2 \theta_W \right) M_Z^2 \cos 2 \beta +L_5 D_L
	&
	M_{e_5}  \ X_{e_5}
	\\
	M_{e_5} \ X_{e_5}
	&
	M_{e_5}^2 \ + \ M_{\tilde{e}_5}^2  +  \sin^2 \theta_W M_Z^2 \cos 2 \beta - L_5 D_L
\end{array}
\right), \nonumber \\
\end{eqnarray}
with $X_{e_5}=A_{e_5} - \mu \tan \beta$ and $L_5 = -(L_4+3)$.
Notice that in order to have the mass matrices for squarks in the MSSM one has to set $D_B=0$.
Knowing the sfermion spectrum we are ready to discuss the predictions for the Higgs mass.
\section{Contributions to $h \rightarrow \gamma\gamma$ and $h \rightarrow g g $}
The formulas presented in this section were adopted from Ref. \cite{Djouadi:1996pb}.
The two-photon decay width of a CP-even Higgs particle $h$ can be written as,
\bea
\Gamma(h \rightarrow \gamma\gamma) = \frac{G_F\alpha_{\rm EW}^2M_{h}^3}{128\sqrt2 \pi^3} \bigg|\sum_i A_i(\tau_i)\bigg|^2\ ,
\eea
where $\tau_i = M_\Phi^2/(4 m_i^2)$ with $m_i$ being the mass of the loop particle. The amplitudes are given by,
\bea\label{As}
A_W(\tau_W) &=& g_{\Phi W W} F_1(\tau_W)\ ,\\
A_f(\tau_f) &=& N_c\, Q_f^2\, g_{\Phi f f} F_{1/2}(\tau_f)\ ,
\\
A_{H^\pm}(\tau_{H^\pm}) &=& g_{\Phi H^+ H^-} \frac{M_W^2}{M_{H^\pm}^2}F_0(\tau_{H^\pm})\ ,\\
A_{\chi_i}(\tau_{\chi_i}) &=& g_{\Phi \chi_i^+ \chi_i^-} \frac{M_W}{m_{\chi_i}}F_{1/2}(\tau_{\chi_i})\ ,\\
A_{\tilde{f}_i}(\tau_{\tilde{f}_i}) &=& N_c \,Q_f^2 \,g_{\Phi \tilde{f}_i \tilde{f}_i} \frac{M_Z^2}{m_{\tilde{f}_i}^2}F_0(\tau_{\tilde{f}_i})\ ,
\eea
where $N_c$ is the color factor and $Q_f$ is the electric charge of the fermion/sfermion in units of the proton charge. The functions $F$ are given by,
\bea
F_0(\tau) &=& \frac{\tau-f(\tau)}{\tau^2}\ ,\\
F_{1/2}(\tau)  &=& -\frac{2[\tau+(\tau-1)f(\tau)]}{\tau^2}\ ,\\
F_1(\tau) &=& \frac{2\tau^2 + 3\tau +3(2\tau-1)f(\tau)}{\tau^2}\ ,
\eea
where,
\bea
f(\tau) = \begin{cases} \ \ {\rm arcsin^2\sqrt{\tau}} \hspace{4.25cm} \tau \le 1\ , \\ \ \ -\frac{1}{4}\left[\log\left(\frac{1+\sqrt{1-1/\tau}}{1-\sqrt{1-1/\tau}}\right)-i\pi\right]^2 \hspace{1cm} \ \tau>1\ .\end{cases}
\eea
\\
\noindent
The mixing angle $\alpha$, which we use in subsequent formulas, is expressed by,
\bea
\alpha = \frac{1}{2}\arctan\left[\frac{2\left({\cal M}_{even}^2\right)_{12}}{\left({\cal M}_{even}^2\right)_{11}-\left({\cal M}_{even}^2\right)_{22}}\right]\ , \ \ \ \alpha \in \left(-\frac{\pi}{2},0\right)\ .
\eea
In the decoupling limit ($M_A^2 \gg M_Z^2$) we obtain  $\alpha \rightarrow \beta - \pi/2$.
\vspace{5mm}
\noindent
Values of MSSM couplings in formulas (\ref{As}):
\begin{itemize}
\item[(a)] $W$ boson loop ($g_{H^0WW}=1$ in SM),
\bea
g_{hWW} = \sin\left(\beta-\alpha\right)\ .
\eea
\item[(b)] Fermion loops ($g_{H^0uu}=g_{H^0dd}=1$ in SM),
\bea
g_{huu} = \frac{\cos\alpha}{\sin\beta}\ , \hspace{0.8cm} g_{hdd} = -\frac{\sin\alpha}{\cos\beta}\ .
\eea
\item[(c)] Charged Higgs loops (negligible in the decoupling regime),
\bea
g_{h H^+ H^-} = \sin\left(\beta-\alpha\right)+ \frac{\cos(2\beta)\sin(\alpha+\beta)}{2\cos^2{\theta_W}} + \frac{\epsilon \cos\alpha \cos^2\beta}{2\cos^2\theta_W M_Z^2 \sin\beta}\ .
\eea
\item[(d)] Top squark loops
\bea
g_{h \tilde{t}_1 \tilde{t}_1} &=& -\frac{1}{2}\sin(\alpha+\beta)\left[\cos^2\theta_t-\frac{4}{3}\sin^2\theta_W\cos2\theta_t\right]+\frac{\cos\alpha}{\sin\beta} \frac{m_t^2}{M_Z^2}\nn\\
&&\ + \frac{m_t \sin2\theta_t}{2 M_Z^2}\left[\frac{\cos\alpha}{\sin\beta}A_t+\frac{\sin\alpha}{\sin\beta}\,\mu\right]\ ,\\
g_{h \tilde{t}_2 \tilde{t}_2} &=&
-\frac{1}{2}\sin(\alpha+\beta)\left[\sin^2\theta_t + \frac{4}{3}\sin^2\theta_W\cos2\theta_t\right]+\frac{\cos\alpha}{\sin\beta} \frac{m_t^2}{M_Z^2}\nn\\
&&\ -\frac{m_t \sin2\theta_t}{2 M_Z^2}\left[\frac{\cos\alpha}{\sin\beta}A_t+\frac{\sin\alpha}{\sin\beta}\,\mu\right]\ ,
\eea
where,
\bea\label{thetaf}
\sin2\theta_t  = \frac{2 m_t X_t}{M^2_{\tilde{t}_1}-M^2_{\tilde{t}_2}} \ .
\eea
\item[(e)] Bottom squark loops
\bea
g_{h \tilde{b}_1 \tilde{b}_1} &=& \frac{1}{2}\sin(\alpha+\beta)\left[\cos^2\theta_b-\frac{2}{3}\sin^2\theta_W\cos2\theta_b\right]-\frac{\sin\alpha}{\cos\beta} \frac{m_b^2}{M_Z^2}\nn\\
&&\ + \frac{m_b \sin2\theta_b}{2 M_Z^2}\left[\frac{\sin\alpha}{\cos\beta}A_b+\frac{\cos\alpha}{\cos\beta}\,\mu\right]\ ,\\
g_{h \tilde{b}_2 \tilde{b}_2} &=&
\frac{1}{2}\sin(\alpha+\beta)\left[\sin^2\theta_b + \frac{2}{3}\sin^2\theta_W\cos2\theta_b\right]-\frac{\sin\alpha}{\cos\beta} \frac{m_b^2}{M_Z^2}\nn\\
&&\ -\frac{m_b \sin2\theta_b}{2 M_Z^2}\left[\frac{\sin\alpha}{\cos\beta}A_b+\frac{\cos\alpha}{\cos\beta}\,\mu\right]\ ,
\eea
where,
\bea\label{thetaf}
\sin2\theta_b  = \frac{2 m_b X_b}{M^2_{\tilde{b}_1}-M^2_{\tilde{b}_2}} \ .
\eea
\item[(f)] Fourth family selectron loops
\bea
g_{h \tilde{e}_{4}^{1} \tilde{e}_{4}^{1}} &=& \frac{1}{2}\sin(\alpha+\beta)\left[\cos^2\theta_{e_4}-2\sin^2\theta_W\cos2\theta_{e_4}\right]-\frac{\sin\alpha}{\cos\beta} \frac{m_{e_4}^2}{M_Z^2}\nn\\
&& \ + \frac{m_{e_4} \sin2\theta_{e_4}}{2 M_Z^2}\left[\frac{\sin\alpha}{\cos\beta}A_{e_4}+\frac{\cos\alpha}{\cos\beta}\,\mu\right]\ ,\\
g_{h \tilde{e}_{4}^{2} \tilde{e}_{4}^{2}} &=& \frac{1}{2}\sin(\alpha+\beta)\left[\sin^2\theta_{e_4}+2\sin^2\theta_W\cos2\theta_{e_4}\right]-\frac{\sin\alpha}{\cos\beta} \frac{m_{e_4}^2}{M_Z^2}\nn\\
&& \ -\frac{m_{e_4} \sin2\theta_{e_4}}{2 M_Z^2}\left[\frac{\sin\alpha}{\cos\beta}A_{e_4}+\frac{\cos\alpha}{\cos\beta}\,\mu\right]\ ,
\eea
where,
\bea
\sin2\theta_{e_4}  = \frac{2 m_{e_4} X_{e_4}}{M^2_{\tilde{e}_{4}^{1}}-M^2_{\tilde{e}_{4}^{2}}}\ .
\eea
\item[(g)] Fifth family selectron loops
\bea
g_{h \tilde{e}_{5}^{1} \tilde{e}_{5}^{1}} &=& -\frac{1}{2}\sin(\alpha+\beta)\left[\cos^2\theta_{e_5}-2\sin^2\theta_W\cos2\theta_{e_5}\right]+\frac{\cos\alpha}{\sin\beta} \frac{m_{e_5}^2}{M_Z^2}\nn\\
&& \ +\frac{m_{e_5} \sin2\theta_{e_5}}{2 M_Z^2}\left[\frac{\cos\alpha}{\sin\beta}A_{e_5}+\frac{\sin\alpha}{\sin\beta}\,\mu\right]\ ,\\
g_{h \tilde{e}_{5}^{2} \tilde{e}_{5}^{2}} &=& -\frac{1}{2}\sin(\alpha+\beta)\left[\sin^2\theta_{e_5}+2\sin^2\theta_W\cos2\theta_{e_5}\right]+\frac{\cos\alpha}{\sin\beta} \frac{m_{e_5}^2}{M_Z^2}\nn\\
&& \ -\frac{m_{e_5} \sin2\theta_{e_5}}{2 M_Z^2}\left[\frac{\cos\alpha}{\sin\beta}A_{e_5}+\frac{\sin\alpha}{\sin\beta}\,\mu\right]\ ,
\eea
where,
\bea
\sin2\theta_{e_5}  = \frac{2 m_{e_5} X_{e_5}}{M^2_{\tilde{e}_{5}^{1}}-M^2_{\tilde{e}_{5}^{2}}}\ .
\eea
\item[(h)] Chargino loops,
\bea
g_{h \chi_1^+ \chi_1^-} &=& \sqrt2 \left(-\cos\alpha \cos\theta^+ \sin\theta^- +\sin\alpha \sin\theta^+ \cos\theta^-\right)\ ,\\
g_{h \chi_2^+ \chi_2^-} &=& -\varepsilon\sqrt2 \left(-\cos\alpha \cos\theta^- \sin\theta^+ +\sin\alpha \sin\theta^- \cos\theta^+\right)\ ,
\eea
where $M_2$ is the gaugino mass parameter, the function $\varepsilon={\rm sign}(\mu M_2 - M_W^2 \sin2\beta)$, and $\theta^\pm$ can be determined from,
\bea
\tan2\theta^+ &=& \frac{2\sqrt2M_W(M_2\cos\beta+\mu \sin\beta)}{M_2^2-\mu^2-2M_W^2\cos2\beta}\ ,\\
\tan2\theta^- &=& \frac{2\sqrt2M_W(M_2\sin\beta+\mu \cos\beta)}{M_2^2-\mu^2+2M_W^2\cos2\beta}\ .
\eea
Chargino masses,
\bea
m^2_{\chi_{1,2}}=&&\frac{1}{2}\bigg[M_2^2+\mu^2+2M_W^2\nn\\
&&\mp \sqrt{(M_2^2-\mu^2)^2 +4M_W^4 \cos^22\beta +4 M_W^2(M_2^2+\mu^2+2 M_2 \mu \sin 2\beta)}\bigg]\ .
\eea
\end{itemize}
The formula below was adopted from Ref. \cite{Djouadi:2005gi}. The two-gluon decay width of a CP-even Higgs particle $h$ is given by,
\bea
\Gamma(h \rightarrow g\, g) = \frac{9G_F\alpha_s^2M_{h}^3}{576\sqrt2 \pi^3} \bigg|\sum_q A_q(\tau_q)+\sum_i A_{\tilde{f}_i}(\tau_{\tilde{f}_i})\bigg|^2\ ,
\eea
where $\tau_i = M_\Phi^2/(4 m_i^2)$ with $m_i$ being the mass of the loop particle and,
\bea
A_q(\tau_q) &=& g_{h q q} F_{1/2}(\tau_q)\ ,
\eea
therefore, it is expressed in terms of quantities we already know how to calculate from the $h \rightarrow \gamma\gamma$ case.
\section{RGEs for the Yukawa Couplings}
The RGEs in the Yukawa sector below the SUSY scale are modified by the new leptons. 
Ignoring the lepton and baryon number gauge contributions which are small below the SUSY scale, 
the Yukawa RGEs below the SUSY scale are
\begin{eqnarray}
	16 \pi^2 \frac{d h_t}{dt} &= & h_t
	\left[
		\frac{9}{2} h_t^2 + \frac{3}{2} h_b^2 + h_\tau^2+ h_{e_4}^2+ h_{e_5}^2 + h_{\nu_4}^2 + h_{\nu_5}^2
		- 4 \pi
		\left(
			\frac{17}{20} \alpha_1 + \frac{9}{4} \alpha_2 + 8 \alpha_3
		\right)
	\right],
\\
	16 \pi^2 \frac{d h_b}{dt} & = & h_b
	\left[
		\frac{9}{2} h_b^2 + \frac{3}{2} h_t^2 + h_\tau^2+ h_{e_4}^2+ h_{e_5}^2 + h_{\nu_4}^2 + h_{\nu_5}^2
		- 4 \pi
		\left(
			\frac{1}{4} \alpha_1 + \frac{9}{4} \alpha_2 + 8 \alpha_3
		\right)
	\right],
\\
	16 \pi^2 \frac{d h_\tau}{dt} & = & h_\tau
	\left[
		\frac{5}{2}h_\tau^2+3 h_b^2 + 3 h_t^2 +  h_{e_4}^2+ h_{e_5}^2 + h_{\nu_4}^2 + h_{\nu_5}^2
		- 4 \pi
		\left(
			\frac{9}{4} \alpha_1 + \frac{9}{4} \alpha_2
		\right)
	\right],
\\
	16 \pi^2 \frac{d h_{\nu_4}}{dt} & = & h_{\nu_4}
	\left[
		\frac{5}{2}h_{\nu_4}^2 - \frac{1}{2} h_{e_4}^2+3 h_b^2 + 3 h_t^2+ h_\tau^2 + h_{e_5}^2 + h_{\nu_5}^2
		- 4 \pi
		\left(
			\frac{3}{4} \alpha_1 + \frac{9}{4} \alpha_2
		\right)
	\right],
\\
	16 \pi^2 \frac{d h_{e_4}}{dt} & = & h_{e_4}
	\left[
		\frac{5}{2}h_{e_4}^2 - \frac{1}{2} h_{\nu_4}^2+3 h_b^2 + 3 h_t^2+ h_\tau^2 + h_{e_5}^2 + h_{\nu_5}^2
		- 4 \pi
		\left(
			\frac{9}{4} \alpha_1 + \frac{9}{4} \alpha_2
		\right)
	\right],
\\
	16 \pi^2 \frac{d h_{\nu_5}}{dt} & = & h_{\nu_5}
	\left[
		\frac{5}{2}h_{\nu_5}^2 - \frac{1}{2} h_{e_5}^2+3 h_b^2 + 3 h_t^2+ h_\tau^2 + h_{e_4}^2 + h_{\nu_4}^2
		- 4 \pi
		\left(
			\frac{3}{4} \alpha_1 + \frac{9}{4} \alpha_2
		\right)
	\right],
\\
	16 \pi^2 \frac{d h_{e_5}}{dt} & = & h_{e_5}
	\left[
		\frac{5}{2}h_{e_5}^2 - \frac{1}{2} h_{\nu_5}^2+3 h_b^2 + 3 h_t^2+ h_\tau^2 + h_{e_4}^2 + h_{\nu_4}^2
		- 4 \pi
		\left(
			\frac{9}{4} \alpha_1 + \frac{9}{4} \alpha_2
		\right)
	\right].
\end{eqnarray}
It has been assumed throughout this paper that the Higgs contribution to the masses of the new quarks is negligible translating into small 
Yukawa couplings that will not greatly affect the running of other Yukawa couplings. They are therefore neglected in the following 
Yukawa RGEs above the SUSY scale:
\begin{eqnarray}
	16 \pi^2 \frac{d Y_t}{dt} &=& Y_t
	\left[
		6 Y_t^2 + Y_b^2 + Y_{\nu_4}^2+ Y_{e_5}^2
		- 4 \pi
		\left(
			\frac{13}{15} \alpha_1 + 3 \alpha_2 + \frac{16}{3} \alpha_3 + \frac{4}{9} N_B \alpha_B
		\right)
	\right],
\\
	16 \pi^2 \frac{d Y_b}{dt} &= &Y_b
	\left[
		6 Y_b^2 + Y_t^2 + Y_\tau^2 + Y_{e_4}^2+ Y_{\nu_5}^2
		- 4 \pi
		\left(
			\frac{7}{15} \alpha_1 + 3 \alpha_2 + \frac{16}{3} \alpha_3 + \frac{4}{9} N_B \alpha_B
		\right)
	\right],
\\
	16 \pi^2 \frac{d Y_\tau}{dt} &=& Y_\tau
	\left[
		4 Y_\tau^2 + 3 Y_b^2 + Y_{e_4}^2+ Y_{\nu_5}^2
		- 4 \pi
		\left(
			\frac{9}{5} \alpha_1 + 3 \alpha_2 + 4 N_L \alpha_L
		\right)
	\right],
\\
	16 \pi^2 \frac{d Y_{e_4}}{dt} &=& Y_{e_4}
	\left[
		4 Y_{e_4}^2 + 3 Y_b^2 + Y_{\tau}^2 + Y_{\nu_4}^2 + Y_{\nu_5}^2
		- 4 \pi
		\left(
			\frac{9}{5} \alpha_1 + 3 \alpha_2 + 4 L_4^2 N_L \alpha_L
		\right)
	\right],
\\
	16 \pi^2 \frac{d Y_{\nu_4}}{dt} &=& Y_{\nu_4}
	\left[
		4 Y_{\nu_4}^2 + 3 Y_t^2+ Y_{e_4}^2+ Y_{e_5}^2
		- 4 \pi
		\left(
			\frac{3}{5} \alpha_1 + 3 \alpha_2 + 4 L_4^2 N_L \alpha_L
		\right)
	\right],
\\
	16 \pi^2 \frac{d Y_{e_5}}{dt} &=& Y_{e_5}
	\left[
		4 Y_{e_5}^2 + 3 Y_t^2 + Y_{\nu_4}^2+ Y_{\nu_5}^2
		- 4 \pi
		\left(
			\frac{9}{5} \alpha_1 + 3 \alpha_2 + 4 N_L (L_4+3)^2 \alpha_L
		\right)
	\right],
\\
	16 \pi^2 \frac{d Y_{\nu_5}}{dt} &=& Y_{\nu_5}
	\left[
		4 Y_{\nu_5}^2 + 3 Y_b^2 + Y_\tau^2 + Y_{e_4}^2+ Y_{e_5}^2
		- 4 \pi
		\left(
			\frac{3}{5} \alpha_1 + 3 \alpha_2 + 4 N_L (L_4+3)^2 \alpha_L
		\right)
	\right]. 
\end{eqnarray}
We remind the reader that $N_B$ and $N_L$ the normalization factors for $U(1)_B$ and $U(1)_L$ and has been both chosen to be half for the numerical work in this paper for simplicity.


\begin{thebibliography}{000}

\bibitem{SUSY-CMS}
\textit{Supersymmetry Searches at CMS},
http://cms.web.cern.ch/org/cms-papers-and-results

\bibitem{SUSY-ATLAS}
\textit{Supersymmetry Searches at ATLAS},
https://twiki.cern.ch/twiki/bin/view/AtlasPublic/SupersymmetryPublicResults

\bibitem{BLMSSM}
  P.~Fileviez P\'erez and M.~B.~Wise,
  \textit{Breaking local baryon and lepton number at the TeV scale},
  JHEP {\bf 1108}, 068 (2011)
  [arXiv:1106.0343 [hep-ph]].
  
\bibitem{FileviezPerez:2010gw}
  P.~Fileviez Perez and M.~B.~Wise,
  \textit{Baryon and lepton number as local gauge symmetries},
  Phys.\ Rev.\ D {\bf 82} (2010) 011901
   [Erratum-ibid.\ D {\bf 82} (2010) 079901]
  [arXiv:1002.1754 [hep-ph]].
\\
See also:
  P.~Fileviez Perez and M.~B.~Wise,
  \textit{Low Energy Supersymmetry with Baryon and Lepton Number Gauged},
  Phys.\ Rev.\ D {\bf 84} (2011) 055015
  [arXiv:1105.3190 [hep-ph]].
  
\bibitem{Dulaney:2010dj}
  T.~R.~Dulaney, P.~Fileviez Perez and M.~B.~Wise,
  \textit{Dark Matter, Baryon Asymmetry, and Spontaneous B and L Breaking},
  Phys.\ Rev.\ D {\bf 83} (2011) 023520
  [arXiv:1005.0617 [hep-ph]].
  
\bibitem{Campbell:1990fa}
  B.~A.~Campbell, S.~Davidson, J.~R.~Ellis, K.~A.~Olive,
  \textit{Cosmological baryon asymmetry constraints on extensions of the standard model},
  Phys.\ Lett.\  {\bf B256 } (1991)  457.

\bibitem{Campbell:1991at}
  B.~A.~Campbell, S.~Davidson, J.~R.~Ellis, K.~A.~Olive,
  \textit{On B+L violation in the laboratory in the light of cosmological and astrophysical constraints},
  Astropart.\ Phys.\  {\bf 1 } (1992)  77-98.

\bibitem{Fischler:1990gn}
  W.~Fischler, G.~F.~Giudice, R.~G.~Leigh, S.~Paban,
  \textit{Constraints on the baryogenesis scale from neutrino masses},
  Phys.\ Lett.\  {\bf B258 } (1991)  45-48.

\bibitem{Dreiner:1992vm}
  H.~K.~Dreiner, G.~G.~Ross,
  \textit{Sphaleron erasure of primordial baryogenesis},
  Nucl.\ Phys.\  {\bf B410 } (1993)  188-216.
  [hep-ph/9207221].

\bibitem{Farrar:1978xj}
  G.~R.~Farrar, P.~Fayet,
  \textit{Phenomenology of the production, decay, and detection of new hadronic states associated with supersymmetry},
  Phys.\ Lett.\  {\bf B76 } (1978)  575-579.

\bibitem{Kuhn:1983sc}
  J.~H.~Kuhn, S.~Ono,
  \textit{Production and decay of gluino - gluino bound states},
  Phys.\ Lett.\  {\bf B142 } (1984)  436.

\bibitem{Goldman:1984mj}
  J.~T.~Goldman, H.~Haber,
  \textit{Gluinonium: The hydrogen atom of supersymmetry},
  Physica {\bf 15D } (1985)  181-196.

\bibitem{Bigi:1991mi}
  I.~I.~Y.~Bigi, V.~S.~Fadin, V.~A.~Khoze,
  \textit{Stop near threshold},
  Nucl.\ Phys.\  {\bf B377 } (1992)  461-479.

\bibitem{Chikovani:1996bk}
  E.~Chikovani, V.~Kartvelishvili, R.~Shanidze, G.~Shaw,
  \textit{Bound states of two gluinos at the Tevatron and CERN LHC},
  Phys.\ Rev.\  {\bf D53 } (1996)  6653-6657.
  [hep-ph/9602249].

\bibitem{Cheung:2004ad}
  K.~Cheung, W.~-Y.~Keung,
  \textit{Split supersymmetry, stable gluino, and gluinonium},
  Phys.\ Rev.\  {\bf D71 } (2005)  015015.
  [hep-ph/0408335].

\bibitem{Drees:1990yw}
  M.~Drees, X.~Tata,
  \textit{Signals for heavy exotics at hadron colliders and supercolliders},
  Phys.\ Lett.\  {\bf B252 } (1990)  695-702.

\bibitem{Feng:1997zr}
  J.~L.~Feng, T.~Moroi,
  \textit{Tevatron signatures of longlived charged sleptons in gauge mediated supersymmetry breaking models},
  Phys.\ Rev.\  {\bf D58 } (1998)  035001.
  [hep-ph/9712499].

\bibitem{Litos}
Michael D. Litos, 
\textit{A search for dinucleon decay into kaons using
the SK water cherenkov detector}, Ph.D. Thesis, Boston
University, 2010.

\bibitem{Goity:1994dq}
  J.~L.~Goity and M.~Sher,
  \textit{Bounds on delta B = 1 couplings in the supersymmetric standard model},
  Phys.\ Lett.\ B {\bf 346} (1995) 69
   [Erratum-ibid.\ B {\bf 385} (1996) 500]
  [hep-ph/9412208].
  

\bibitem{Collaboration:2012sk}
  ATLAS Collaboration,
  \textit{Search for the standard model Higgs boson in the diphoton decay channel with 4.9 fb-1 of pp collisions at sqrt(s)=7 TeV with ATLAS},
  arXiv:1202.1414 [hep-ex].

\bibitem{Collaboration:2012si}
  ATLAS Collaboration,
  \textit{Combined search for the standard model Higgs boson using up to 4.9 fb-1 of pp collision data at sqrt(s) = 7 TeV with the ATLAS detector at the LHC},
  arXiv:1202.1408 [hep-ex].


\bibitem{Collaboration:2012tx}
  CMS Collaboration,
  \textit{Combined results of searches for the standard model Higgs boson in pp collisions at sqrt(s) = 7 TeV},
  arXiv:1202.1488 [hep-ex].

\bibitem{Collaboration:2012tw}
  CMS Collaboration,
  \textit{Search for the standard model Higgs boson decaying into two photons in pp collisions at sqrt(s)=7 TeV},
  arXiv:1202.1487 [hep-ex].
  
\bibitem{CMSnew}
https://cdsweb.cern.ch/record/1429928/files/HIG-12-008-pas.pdf


\bibitem{FeynHiggs}
  S.~Heinemeyer, W.~Hollik and G.~Weiglein,
  \textit{FeynHiggs: A Program for the calculation of the masses of the neutral CP even Higgs bosons in the MSSM},
  Comput.\ Phys.\ Commun.\  {\bf 124} (2000) 76
  [hep-ph/9812320];
  \\
  S.~Heinemeyer, W.~Hollik and G.~Weiglein,
  \textit{The Masses of the neutral CP - even Higgs bosons in the MSSM: Accurate analysis at the two loop level},
  Eur.\ Phys.\ J.\ C {\bf 9} (1999) 343
  [hep-ph/9812472];
  \\
  G.~Degrassi, S.~Heinemeyer, W.~Hollik, P.~Slavich and G.~Weiglein,
  \textit{Towards high precision predictions for the MSSM Higgs sector},
  Eur.\ Phys.\ J.\ C {\bf 28} (2003) 133
  [hep-ph/0212020];
  \\
  M.~Frank, T.~Hahn, S.~Heinemeyer, W.~Hollik, H.~Rzehak and G.~Weiglein,
  \textit{The Higgs Boson Masses and Mixings of the Complex MSSM in the Feynman-Diagrammatic Approach},
  JHEP {\bf 0702} (2007) 047
  [hep-ph/0611326].
  
\bibitem{FileviezPerez:2012iw}
  P.~Fileviez Perez,
  \textit{SUSY Spectrum and the Higgs Mass in the BLMSSM},
  arXiv:1201.1501 [hep-ph], to appear in Physics Letters B.
  
\bibitem{Djouadi:1996pb}
  A.~Djouadi, V.~Driesen, W.~Hollik and J.~I.~Illana,
  \textit{The coupling of the lightest SUSY Higgs boson to two photons in the decoupling regime},
  Eur.\ Phys.\ J.\ C {\bf 1}, 149 (1998)
  [hep-ph/9612362].
  
\bibitem{Ishiwata:2011hr}
  K.~Ishiwata and M.~B.~Wise,
  \textit{Higgs Properties and Fourth Generation Leptons},
  Phys.\ Rev.\ D {\bf 84} (2011) 055025
  [arXiv:1107.1490 [hep-ph]].
  
\bibitem{Spira:1997dg}
  M.~Spira,
  \textit{QCD effects in Higgs physics},
  Fortsch.\ Phys.\  {\bf 46} (1998) 203
  [hep-ph/9705337].
  
\bibitem{Djouadi:2005gi}
  A.~Djouadi,
  \textit{The anatomy of electro-weak symmetry breaking. I: The Higgs boson in the standard model},
  Phys.\ Rept.\  {\bf 457}, 1 (2008)
  [hep-ph/0503172].

\end{thebibliography}
\end{document}